\documentclass[journal]{IEEEtran}
\newif\ifhavebib
\usepackage{blindtext}
\usepackage{graphicx}
\usepackage{graphicx,subfigure}
\usepackage{array}
\usepackage{url}
\usepackage{algorithmic}
\usepackage[cmex10]{amsmath}
\usepackage{ifpdf}
\usepackage{amssymb,amsmath}
\usepackage{empheq}
\usepackage{stfloats}
\usepackage{textcomp}
\usepackage{cite}
\usepackage{multirow}
\usepackage{diagbox}
\usepackage[ruled,vlined,lined,ruled,boxed]{algorithm2e}
\usepackage{fixltx2e}
\usepackage[dvipsnames,usenames]{color}
\usepackage[normalem]{ulem}
\usepackage{amsthm}

\usepackage{graphicx}
\usepackage{ifpdf}
\usepackage{cite}
\usepackage{enumerate}
\usepackage{enumitem}

\usepackage{array}
\usepackage{mdwmath}
\usepackage{mdwtab}
\usepackage{eqparbox}
\usepackage{bm}
\usepackage{setspace}
\usepackage{booktabs}
\usepackage[subnum]{cases}
\usepackage{rotating}
\definecolor{Red}{rgb}{1,0,0}
\definecolor{Blue}{rgb}{0,0,1}

\usepackage[colorlinks=true, urlcolor=black,  linkcolor=black, citecolor=black]{hyperref}
\theoremstyle{definition}

\theoremstyle{remark}
\newtheorem{remark}{Remark}

\newtheorem{fact}{Fact}
\theoremstyle{definition}

\DeclarePairedDelimiterX{\infdivx}[2]{(}{)}{%
	#1\;\delimsize\|\;#2%
}

\DeclarePairedDelimiter{\norm}{\lVert}{\rVert}
\DeclarePairedDelimiter{\abs}{\lvert}{\rvert}
\DeclareMathOperator*{\argmax}{argmax}

\def\Re{\mathop{\rm Re}\nolimits}%
\newcommand{\av}{{\bf a}}

\newcommand{\Cv}{{\bf C}}

\newcommand{\Xv}{{\bf X}}
\newcommand{\Yv}{{\bf Y}}
\newcommand{\Zv}{{\bf Z}}
\newcommand{\Uv}{{\bf U}}
\newcommand{\Fv}{{\bf F}}
\newcommand{\Vv}{{\bf V}}

\newcommand{\Qv}{{\bf Q}}

\newcommand{\Av}{{\bf A}}

\newcommand{\Hv}{{\bf H}}

\newcommand{\Sv}{{\bf S}}
\newcommand{\Nv}{{\bf N}}

\newcommand{\xv}{{\bf x}}

\newcommand{\yv}{{\bf y}}

\newcommand{\hv}{{\bf h}}
\newcommand{\sv}{{\bf s}}
\newcommand{\nv}{{\bf n}}

\newcommand{\Wv}{{\bf W}}

\newcommand{\varphiv}{\boldsymbol \varphi}
\newcommand{\thetav}{\boldsymbol \theta}
\newcommand{\varthetav}{\boldsymbol \vartheta}

\newcommand{\Xiv}{\boldsymbol \Xi}

\newcommand{\ph}{{\hat{p}}}
\newcommand{\qh}{{\hat{q}}}
\newcommand{\sh}{{\hat{s}}}
\newcommand{\uh}{{\hat{u}}}

\newcommand{\wh}{{\hat{w}}}
\newcommand{\xh}{{\hat{x}}}

\newcommand{\rh}{{\hat{r}}}
\newcommand{\zh}{{\hat{z}}}

\newcommand{\betah}{\hat \beta}
\newcommand{\gammah}{\hat \gamma}
\newcommand{\zetah}{\hat \zeta}

\newcommand{\Hvh}{{\hat{\bf H}}}
\newcommand{\Svh}{{\hat{\bf S}}}

\newcommand{\xvh}{{\hat{\bf x}}}
\newcommand{\Xvh}{{\hat{\bf X}}}
\newcommand{\Wvh}{{\hat{\bf W}}}
\newcommand{\Zvh}{{\hat{\bf Z}}}

\newcommand{\Yvt}{{\tilde{\Yv}}}
\newcommand{\Svt}{{\tilde{\Sv}}}
\newcommand{\Nvt}{{\tilde{\Nv}}}

\newcommand{\hvt}{{\tilde{\hv}}}

\def\a{\alpha}

\DeclareMathOperator\E{E}

\def\E{\mathbb{E}}

\newcommand{\Norm}{\mathcal{N}}
\newcommand{\CN}{\mathcal{CN}}

\newcommand\eg{e.g.,\xspace}
\newcommand\ie{i.e.,\xspace}
\def\textiid{i.i.d.\@\xspace}
\newcommand\iid{\ifmmode\text{ i.i.d. } \else \textiid \fi}

\newcommand{\Complex}{\mathbb{C}}
\newcommand{\Real}{\mathbb{R}}

\begin{document}
\setitemize{listparindent=\parindent,partopsep=0pt,topsep=-0.25ex}
\setenumerate{fullwidth,itemindent=\parindent,listparindent=\parindent,itemsep=0ex,partopsep=0pt,parsep=0ex}
\havebibtrue
\title{Super-Resolution Blind Channel-and-Signal\\ Estimation for Massive MIMO with\\ One-Dimensional Antenna Array}
\author{Sep. 21st, 2018}
\author{Hang~Liu, Xiaojun~Yuan,~\IEEEmembership{Senior Member,~IEEE,}
	and~Ying~Jun (Angela)~Zhang,~\IEEEmembership{Senior~Member,~IEEE}
		\thanks{This work was supported in part by General Research Funding (Project number 14209414, 14208107) from the Research Grants Council of Hong Kong. The work of X. Yuan was supported in part by Guangdong Provincial Key Area of Research and Development Program of China under Grant: 2018B010114001. This work was presented in part at IEEE International Conference on Communications, Shanghai, P.R. China, May 2019 \cite{ICC19}.}
		\thanks{H. Liu and Y. J. Zhang are with the Department of Information Engineering, The Chinese University of Hong Kong, Shatin, New Territories, Hong Kong (e-mail: lh117@ie.cuhk.edu.hk; yjzhang@ie.cuhk.edu.hk).}
		\thanks{X. Yuan is with the Center for Intelligent Networking and Communications, the University of Electronic Science and Technology of China, Chengdu, China (e-mail: xjyuan@uestc.edu.cn).}
}
\maketitle
\begin{abstract}
In this paper, we study blind channel-and-signal estimation by exploiting the burst-sparse structure of angular-domain propagation channels in massive MIMO systems. The state-of-the-art approach utilizes the structured channel sparsity by sampling the angular-domain channel representation with a uniform angle-sampling grid, a.k.a. virtual channel representation. However, this approach is only applicable to uniform linear arrays and may cause a substantial performance loss due to the mismatch between the virtual representation and the true angle information. To tackle these challenges, we propose a sparse channel representation with a super-resolution sampling grid and a hidden Markovian support. Based on this, we develop a novel approximate inference based blind estimation algorithm to estimate the channel and the user signals simultaneously, with emphasis on the adoption of the expectation-maximization method to learn the angle information. Furthermore, we demonstrate the low-complexity implementation of our algorithm, making use of factor graph and message passing principles to compute the marginal posteriors. Numerical results show that our proposed method significantly reduces the estimation error compared to the state-of-the-art approach under various settings, which verifies the efficiency and robustness of our method.
\end{abstract}
\IEEEpeerreviewmaketitle
\section{Introduction}
\IEEEPARstart{M}{ultiuser} massive multi-input multi-output (MIMO) has attracted intensive research interests since it provides remarkable improvements on system capacity and reliability\cite{MIMO_over1, MIMO_over3, MIMO_over4, MIMO_over5}. As one of the key obstacles to utilizing the high array gain of a massive MIMO system, the acquisition of the channel state information (CSI) becomes challenging due to increased  channel dimensions and fast channel variations \cite{MIMO_chanllenge}. Many studies have been conducted to design reliable techniques for channel acquisition. 
For example, a conventional training-based approach estimates the channel coefficients by using orthogonal pilot sequences \cite{MIMO_pilot3,MIMO_pilot4,MIMO_chanllenge2}. Joint channel and signal estimation can further improve the system performance since the estimated signals can be used as ``soft pilots” to enhance the channel estimation accuracy \cite{MIMO_pilot7,MIMO_bayes}. However, the training overhead may become unaffordable in a large-dimension system  due to the constrained length of the channel coherence time \cite{MIMO_chanllenge3}. Moreover, training-based methods give rise to the pilot contamination problem, for the reason that the number of available orthogonal pilot sequences cannot exceed the sequence length \cite{MIMO_contamination}.

In contrast to training-based methods, another line of research, namely blind or semi-blind channel estimation, aims to estimate the channels without or with less help of the training processes. In \cite{MIMO_blind1,MIMO_EVD,MIMO_Neumann}, a particular class of algorithms under this category was proposed to rely on a subspace partition of the received signals, by assuming the asymptotic orthogonality of user channels for extremely large MIMO systems. However, subspace-based methods suffer severe estimation inaccuracy since the number of antennas at the receiver is always limited to be finite in practice. More importantly, these methods require very long coherence duration to generate stable channel estimates, which is usually difficult to realize in fast time-varying scenarios.

Recently, experimental studies have evidenced the burst-sparse structure in the angular domain of the physical channels in massive MIMO systems, thanks to the limited number of scatterers in the propagation environment \cite{Channel_Geometry,Channel_3DmmWave}. Inspired by this, recent studies exploited the sparsity of the massive MIMO channel \cite{MIMO_pilot5,MIMO_pilot6} or {the low-rankness of the channel covariance matrix \cite{Channel_formular,MIMO_TSPrecent} in the design of training-based channel estimation schemes}. The burst-sparse structure of the channel was modeled using Markovian priors to further improve the channel estimation performance \cite{MIMO_TurboCS,MIMO_LiuAN}. Alternatively, a blind channel estimation method based on channel sparsity was proposed in \cite{MIMO_Amine}, where the subspace partition with $\ell_1$ regularization was developed to force the sparsity of the channel matrix under the discrete Fourier transform (DFT) angular basis. 
 
To exploit the channel sparsity more efficiently, the authors in \cite{MIMO_zhangjianwen} developed a blind massive MIMO channel-and-signal estimation scheme that simultaneously estimates the channel and detects the transmitted signal from the received signal. It was shown in \cite{MIMO_zhangjianwen} that the channel sparsity leads to a fundamental performance gain on the degrees of freedom of the massive MIMO system. The authors in \cite{MIMO_zhangjianwen} also proposed a modified bilinear generalized approximate message passing (BiG-AMP) algorithm \cite{MP_BIGAMP1}, termed projection-based BiG-AMP (Pro-BiG-AMP), to efficiently factorize the sparse channel matrix and the signal matrix.


The existing sparsity-learning based blind channel estimation methods \cite{MIMO_Amine,MIMO_zhangjianwen} assume half-wavelength uniform linear antenna arrays (ULAs). Then, with a uniform sampling grid and a DFT basis, the virtual channel representation model \cite{Channel_virtual_representation,Channel_ExperimentalVR} is employed to characterize the angular-domain sparsity of the massive MIMO channel. However, the DFT-based virtual channel representation is not applicable to ULAs with a non-uniform sampling grid, letting alone antenna arrays with arbitrary geometry. More importantly, the channel estimation accuracy of all DFT-based methods is severely compromised even for ULAs, not only due to the leakage of energy in the DFT basis, but also because the DFT basis is not adaptable to efficiently exploit the burst-sparse structure of the physical channels. To tackle the defects of the existing blind estimation methods, we model the massive MIMO channel using the non-uniform angle-sampling \cite{MIMO_Ding} and the off-grid representation \cite{MIMO_DOA1,OFFGRID_1,OFFGRID_2,OFFGRID_3}. Based on this, we formulate the blind channel estimation problem as an affine matrix factorization (AMF) problem and propose an efficient super-resolution blind channel-and-signal estimation algorithm based on approximate inference and expectation-maximization (EM) principles. The main contributions of this paper are summarized as follows. 
\begin{itemize}
	\item 	\emph{Non-uniform angle-sampling and sparse representation with Markovian support for blind channel-and-signal estimation problem}:
	
	Unlike the existing blind estimation methods that adopt the DFT basis representation, we unfix the angle-of-arrival (AoA) sampling grid and employ an off-grid channel model in massive MIMO. Although the non-DFT sampling basis and the off-grid model have been employed previously in \cite{MIMO_Ding, MIMO_DOA1}, this is the first work to adapt the idea to the joint channel-and-signal estimation problem. As such, we formulate the problem as an AMF task with unknown parameters, where the existing off-grid frameworks are not applicable.   
	Following the prior work in \cite{MIMO_TurboCS,MIMO_LiuAN}, we employ a set of Markov chains, one for each user, to model the probability space of the channel support. The framework does not impose any restrictions on angle sampling, and hence is able to avoid the energy leakage problem by achieving a resolution much higher than uniform sampling. Moreover, with the Markovian structure, our proposed framework is able to capture the burst-sparse nature of the massive MIMO channel.
	\item 	\emph{Super-resolution blind channel-and-signal estimation via approximate-inference-based EM}:
	
	We develop a novel blind channel-and-signal estimation method based on EM principles. We show that the exact posterior distributions required by the expectation step (E-step) is difficult to acquire for the bilinear system model with both channel and signal unknown. As such, we propose to utilize approximate inference for the realization of the E-step, hence the name approximate-inference-based EM. Since the proposed solution is designed on top of the off-grid channel representation with the Markovian support, our solution operates with {general one-dimensional array geometry} and overcomes the angle mismatch problem in the DFT-based methods. Furthermore, under the EM framework, angle-tuning and hyper-parameter learning are employed to learn the AoAs automatically without any prior knowledge of the CSI. We also show that the proposed scheme demonstrates a superior performance to the existing super-resolution algorithm in  \cite{MIMO_DOA1} under the joint channel-and-signal estimation framework. 
	\item \emph{Marginal posterior calculation via message passing}:
	
	We present a low-complexity implementation of the proposed algorithm. Specifically, as the EM involves the calculation of marginals, we construct an associated factor graph and apply the message passing principles to achieve this purpose. To further reduce the computation complexity, additional approximations are introduced to particular messages based on the general approximate message passing (AMP) framework \cite{MP_AMP1,MP_GAMP}. Finally, we put forth a simplified message passing algorithm and concrete maximization step (M-step) update rules. Additional guidance for convergence acceleration is provided as well.
\end{itemize}

The remainder of this paper is organized as follows: Section \ref{systemsetup} describes the system model and introduces the off-grid representation. Investigations on the state-of-the-art blind channel estimation algorithm are conducted in this section as well. In Section \ref{derivation}, we derive the proposed blind channel-and-signal estimation algorithm and the associated prior-information learning algorithm. In Section \ref{section4}, we present the marginal posterior calculation scheme based on message passing. Furthermore, Section \ref{Simulation} gives numerical results on the proposed methods. Finally, the paper concludes in Section \ref{conclusions}.

\emph{Notation}: Throughout, we use $\Real$ and $\Complex$ to denote the real and complex number sets, respectively. 
We use regular small letters, bold small letters, and bold capital letters to denote scalars, vectors, and matrices, respectively. We use $x_{ij}$ to denote the entry at the $i$-th row and $j$-th column of matrix $\Xv$. We use $(\cdot)^\star$, $(\cdot)^T$, and $(\cdot)^H$ to denote the conjugate, transpose, and conjugate transpose, respectively. We use $\E[\cdot]$ to denote the expectation operator, $\abs{\cdot}$ to denote the absolute operator, $\norm{\cdot}_p$ to denote the $\ell_p$ norm, $\norm{\cdot}_F$ to denote the Frobenius norm, $\delta(\cdot)$ to denote the Dirac delta function, and $\propto$ to denote equality up to a constant multiplicative factor. We use $\Norm(\cdot;\mu,\sigma^2)$ and $\CN(\cdot;\mu,\sigma^2)$ to denote the real normal and circularly-symmetric normal distributions with mean $\mu$ and variance $\sigma^2$, respectively. Finally, we define $[n] \triangleq \{1,2,3,\cdots,n \}$ for some positive integer $n$.
\section{System Model \label{systemsetup}}

\subsection{Massive MIMO Channel Model}
\label{channelmodel}
Consider a multiuser massive MIMO system with $K$ users equipped with a single antenna and one base station (BS) with $N$ antennas, where $N \gg K \gg 1$. We assume that the channel is block-fading. During coherence time $T$, the uplink channel impulse response from the $k$-th user to the BS can be modeled as \cite{Channel_formular}
\begin{equation}
\label{hk1}
\hv_k(t)=\sum_{i=1}^{L_c(k)} \sum_{j=1}^{L_p(k)} \a_k(i,j) \av\left(\theta_k(i,j)\right) \delta\left(t\!-\!\tau_k(i,j)\right) , \forall t \in [T],
\end{equation}
where $L_c(k)$ and $L_p(k)$ denote the number of scattering clusters and the number of physical paths in each cluster between the $k$-th user and the BS, respectively; $\a_k(i,j)$ denotes the complex-valued channel coefficient of the $j$-th path in the $i$-th cluster for the $k$-th user; $\theta_k(i,j)$ denotes the corresponding azimuth AoA; $\tau_k(i,j)$ denotes the corresponding time delay; and $\av(\theta)$ is the steering vector for receiving a signal, impinging upon the antenna array at angle $\theta$.\footnote{
	In this paper, we restrict our discussion on the {one-dimensional} array geometry with the steering vector only related to the azimuth angle.} In general, $\av(\theta)$ is determined by the geometry of the antenna array at the BS. For instance, if a ULA is deployed at the BS, $\av(\theta)$ is given by
\begin{equation}
\label{ULA}
\av(\theta)=\frac{1}{\sqrt{N}}\left[1,e^{-j\frac{2\pi}{\varrho}d\sin(\theta)},\cdots,e^{-j\frac{2\pi}{\varrho}d(N-1)\sin(\theta)}\right]^T,
\end{equation}
where $\varrho$ denotes the carrier wavelength, and $d$ denotes the distance between any two adjacent antennas. As another example, suppose that the BS is equipped with a lens antenna array (LAA), where the lens antennas are placed on the focal arc of the lens with critical antenna spacing. The steering vector $\av(\theta)$ is given by \eqref{LAA}, where $\text{sinc}(\cdot)$ denotes the nominalized ``\emph{sinc}" function, and $D$ denotes the lens length along the azimuth plane \cite{MIMO_LAA}.
\begin{figure*}[ht]
		\begin{equation}
		\label{LAA}
		\av(\theta)=\left[\text{sinc}\left(-\frac{N-1}{2}-\frac{D}{\varrho}\sin\theta\right),\text{sinc}\left(-\frac{N-3}{2}-\frac{D}{\varrho}\sin\theta\right),\cdots,\text{sinc}\left(\frac{N-1}{2}-\frac{D}{\varrho}\sin\theta\right)\right]^T.
		\end{equation} 
	\hrulefill
\end{figure*}

With the channel impulse response given by \eqref{hk1}, the received  signal at time $t$ can be expressed as
\begin{align}
\label{yk1}
\yv(t)&=\sum_{k=1}^K \hv_k(t)  \ast x_k(t) + \nv(t) \nonumber\\
&= \sum_{k=1}^K\sum_{i=1}^{L_c(k)} \sum_{j=1}^{L_p(k)} \a_k(i,\!j) \av\left(\theta_k(i,\!j)\right) x_k(t\!-\!\tau_k(i,\!j))\!+\!\nv(t),
\end{align}
where $\ast$ denotes the linear convolution operation; $x_k(t) \in \Complex $ is the transmitted symbol from the $k$-th user; and $\nv(t)$ is the channel noise vector. 

Assume that the delay spread for different paths is negligible compared to the reciprocal of signal bandwidth $W$, \ie $|\tau_k(i,j)\!-\!\tau_k(i^\prime,j^\prime)|\!\ll\! 1/W$ for $\forall (i,j),(i^\prime\!,j^\prime\!)$. We further simplify \eqref{yk1} by approximating different delays with the same value, \ie $\tau_k(i,j) \!\approx \!\tau_k$. Therefore, with time synchronization at the receiver side, we can rewrite \eqref{yk1} as
\begin{equation}
\label{yk2}
\yv(t)=  \sum_{k=1}^K\tilde{ \hv}_k x_k(t)+\nv(t),
\end{equation}
where the channel coefficient vector of the $k$-th user is defined as
\begin{equation}
\label{hktilde}
\tilde{ \hv}_k\triangleq\sum_{i=1}^{L_c(k)}\sum_{j=1}^{L_p(k)} \a_k(i,j) \av\left(\theta_k(i,j)\right).
\end{equation}
\subsection{Off-Grid Representation}
\label{signalmodel}
Let $\varthetav=\{\vartheta_l\}^L_{l=1}$ be a given grid that consists of $L$ discrete angular points and covers the AoAs ranging from $-90^\circ$ to $90^\circ$. Denote by $\thetav\triangleq \{\theta_{k}(i,j)\}_{\forall k,i,j}$ the collection of true AoAs.
When $L$ becomes large enough such that $\thetav \!\subseteq\! \varthetav$, \eqref{hktilde} can be represented as
\begin{equation}
\begin{aligned}
\label{hktilde2}
\tilde{ \hv}_k= \sum_{l=1}^L s_{k,l} \av(\vartheta_l)=\Av(\varthetav)\sv_k,
\end{aligned}
\end{equation}
where $\Av(\varthetav)\!\triangleq\! [\av(\vartheta_1),\av(\vartheta_2),\cdots,\av(\vartheta_L)]\!\in\! \Complex^{N\times L}$ is the angular array response for given $\varthetav$, and $\sv_k \triangleq [s_{k,1},s_{k,2},\cdots,s_{k,L}]^T \in \Complex^{L\times 1}$ contains
the corresponding channel coefficients in the angular domain. More specifically, we have
\begin{align}{s_{k,l}\!=\!}
\begin{cases}
\a_k(i,j), &\text{if } \vartheta_l\!=\!\theta_k(i,j) \text{ for some } \theta_k(i,j)\! \in\! \thetav, \\
0, &\text{otherwise.}
\end{cases}
\end{align}

In fact, $\varthetav$ can be viewed as a set of angular resolution bins located at the BS. In practice, since the scattering clusters usually have certain distances from the BS, the subpath associated with a particular scattering cluster will only have a small range of angular spread \cite{Channel_Geometry}. Moreover, since the number of scatterers is usually very small \cite{Channel_Geometry,Channel_3DmmWave}, only a small portion of resolution bins will be occupied \cite{Channel_virtual_representation, Channel_ExperimentalVR}. Therefore, if the selected grid  $\varthetav$ covers the true directions $\thetav$ well, $\sv_k$ will be sparse and the non-zero elements of $\sv_k$ will concentrate in a narrow range around a few positions \cite{MIMO_DOA1}. We refer to this phenomenon as the burst-sparse structure of channel coefficient vectors in the angular domain.
However, the direction mismatch generally exists between $\thetav$ and $\varthetav$, since the true AoAs are difficult to acquire in practice. To address this problem, we propose an automatic angle tuning scheme for $\varthetav$ in Section \ref{derivation}.

For each coherence duration $T$, we denote the transmission signals for the $k$-th user by $\xv_k\!=\!\left[ x_k(1),x_k(2),\cdots,x_k(T)  \right]^T $, and denote the collection of all transmission signals by $\Xv\!=\!\left[\xv_1,\xv_2,\cdots,\xv_K \right]^T\!\in \!\Complex^{K\times T}$.
Then, we represent the received signal by
\begin{equation}
\label{Y1}
\Yv=\Av(\varthetav)\Sv\Xv+\Nv,
\end{equation}
where $\Yv\!=\![\yv(1),\yv(2),\cdots,\yv(T)] \!\in\! \Complex^{N\times T}$ denotes the collection of the received signals; $\Sv\!=\![\sv_1,\sv_2,\cdots, \sv_K] \!\in\! \Complex^{L\times K}$ is the corresponding channel coefficient matrix in the angular domain;  and $\Nv\in \Complex^{N\times T}$ denotes an additive white Gaussian noise (AWGN) matrix with the entries \iid drawn from $\CN(\cdot;0,\sigma^2)$.

Let $P_k$ denote the average transmission power for the $k$-th user in duration $T$. Denote by $P\triangleq \sum_{k=1}^K P_k$ the total average transmission power. For a transmission sequence $\xv_k$, by the definition of $P_k$ we have
\begin{equation}
\label{Xpower}
\frac{1}{T}\E[\xv_k^H\xv_k]=P_k, \text{ for } \forall k \in [K].
\end{equation}
Without loss of generality, we assume $P=K$ in the sequel. Additionally, we assume $P_k=1$ for $\forall k$.\footnote{This assumption is not essential for the derivation of our proposed algorithm. One can readily extend the algorithm to the case of non-uniform power allocation. Here, we assume equal transmission power for simplicity.}


\subsection{Channel Representation with DFT Basis}
\label{DFTdefects}
Here, we discuss a state-of-the-art framework for blindly estimating user signals from \eqref{Y1} and specify its potential limitations. Suppose that the BS is equipped with a ULA. A framework called virtual channel representation \cite{Channel_virtual_representation,Channel_ExperimentalVR} was proposed to resolve the channel by a fixed sampling grid $\varthetav^0$ with length $L=N$. Specifically, $\varthetav^0$ satisfies 
\begin{equation}
\label{VCR}
\frac{d} \varrho\sin(\vartheta^0_n)=\frac{n-1}{N} , \forall n\in [N].
\end{equation}
Substituting \eqref{VCR} into \eqref{ULA}, we see that $\Av(\varthetav^0)$ is the normalized DFT matrix denoted by $\Fv$. 
Thus, we obtain
\begin{equation}
\label{DFTeq}
\Yvt=\Fv^H\Yv= \Svt \Xv+ \Nvt,
\end{equation}
where $\Svt=\Fv^H\Av(\thetav)\Sv\in \Complex^{N\times K} $ is the channel representation under the DFT basis, and the entries of $\Nvt$ are \iid drawn from $\CN(\cdot;0,\sigma^2)$.

Existing blind estimation methods \cite{MIMO_zhangjianwen,MIMO_Amine} aim to recover both the channel $\Svt$ and the signal $\Xv$ from the observation of $\Yvt$ in \eqref{DFTeq} by exploiting the sparsity of $\Svt$. However, this framework suffers from at least three defects:
\begin{itemize}
	\item The DFT matrix can only approximate the array response of a ULA. For a general antenna geometry, we may not be able to project the received signal matrix to the angular domain by a simple unitary transformation.
	\item Since the number of antennas at the BS $N$ is physically constrained to be finite, the DFT-based methods always have performance loss due to the unavoidable AoA mismatch between $\varthetav^0$ and $\thetav$, a.k.a. the energy leakage phenomenon. More importantly, due to the energy leakage, the channel matrix $\Svt$ is not exactly sparse. This may seriously compromise the performance of the channel estimation and signal estimation methods based on the channel sparsity.
	\item As mentioned in Section \ref{signalmodel}, the true AoAs tend to concentrate in a few groups due to a limited number of scattering clusters. 
	The DFT-based methods fail to exploit this burst-sparse structure since they sample the AoA range uniformly.
\end{itemize}

To address the above defects, we develop a novel framework to blindly estimate the channel matrix and the user signals. We tackle the problem directly based on the signal model \eqref{Y1}, without resorting to the DFT-based simplification in  \eqref{DFTeq}. 
{We assume general one-dimensional array geometry at the BS and use a general form of the steering vector $\av(\theta)$ in the derivation. Extensions of the following framework for higher-dimensional antenna arrays are possible but is not the focus of this paper.}
%
%
%
\section{Super-Resolution Blind Channel-and-Signal Estimation\label{derivation}}
In this section, we develop a framework to infer $\Sv$ and $\Xv$ given $\Yv$ in \eqref{Y1}, a.k.a. the AMF problem. Besides, we discuss how to estimate the latent parameters and tune the grid $\varthetav$ in the proposed algorithm. 
\subsection{Probabilistic Model}
\label{mp}
Define $\Wv \triangleq\Sv \Xv \!\in\!  \Complex^{L\times T}$ and $\Zv \triangleq \Av(\varthetav) \Wv \!\in\!  \Complex^{N\times T}$. Under the assumption of AWGN, we have 
\begin{equation}
\label{gout}
p(\Yv|\Zv)=\prod_{n=1}^N\prod_{t=1}^T \CN\left( y_{nt};z_{nt},\sigma^2\right) .
\end{equation}
For simplicity, we assume that all signals are generated independently, \ie
\begin{equation}
\label{gx}
p(\Xv)=\prod_{k=1}^K\prod_{t=1}^T p_{x_{kt}}\left( x_{kt}\right) ,
\end{equation}
where $p_{x_{kt}}(x_{kt})$ is determined by the signal generation model at the transmitter side. Moreover, the hidden sparsity of $\Sv$ motivates us to model the prior distribution of $\Sv$ as
\begin{equation}
\label{gs1}
p(\Sv|\Cv)=\prod_{l=1}^L\prod_{k=1}^K \delta\left( c_{lk}\right) \delta\left( s_{lk}\right)+\delta\left( c_{lk}\!-\!1\right)\CN(s_{lk};0,\varphi_k),
\end{equation}
where the binary state variable $c_{lk}\in \{0,1\}$ for $\forall l,k$ is introduced to indicate whether the corresponding entry of $\Sv$ is $0$ or not. In other words, $\Cv\in \{0,1\}^{L\times K}$ is the support for $\Sv$. Following \cite{MIMO_EVD}, we assign a Gaussian prior distribution with zero mean and distinct variance to each non-zero entry of $\Sv$. Note that here we set $\varphi_k$ to be independent of the antenna grid index $l$.

An intuitive approach to model the support matrix $\Cv$ is to assign an \iid Bernoulli distribution to its entries, \ie
\begin{equation}
\label{gc1}
p\left( c_{lk}=1\right) =\lambda, \forall l,k,
\end{equation}
where $\lambda$ is the Bernoulli parameter. However, as discussed in Section \ref{signalmodel}, the non-zero elements of $\Cv$ can be grouped into a small number of clusters. We characterize such a clustered structure of $\Cv$ using a set of Markov chains as
\begin{equation}
\label{gc2}
p(\Cv)=\prod_{k=1}^K \left( p(c_{1k})\prod_{l=2}^L p\left( c_{lk}|c_{l-1,k}\right)\right).
\end{equation}
The transition probabilities are defined as $p\left( c_{lk}=0|c_{l-1,k}=1\right)\triangleq p_{01}$ and $p\left( c_{lk}=1|c_{l-1,k}=0\right)=p_{01}\lambda/(1-\lambda)\triangleq p_{10}$. These definitions ensure that each Markov chain is consistent with the marginal distribution \eqref{gc1} with $p\left( c_{1k}=1\right)$ set to $\lambda$. Therefore, by marginalizing $c_{lk}$, we obtain
\begin{equation}
\label{gs2}
p_{s_{lk}}(s_{lk})=(1-\lambda)\delta(s_{lk})+\lambda\CN\left(s_{lk};0,\varphi_k \right) .
\end{equation}
\subsection{Inference by Expectation-Maximization \label{bayesian}}
Given the prior distributions, the \emph{maximum a posterior} (MAP) estimator ($\Xvh$, $\Svh$) is given by
 \begin{align}
	\label{map}
	(\Xvh, \Svh) = \argmax_{\Xv,\Sv} \int_{\Cv}p(\Xv,\Sv,\Cv|\Yv).
\end{align}
The computation of \eqref{map} requires the knowledge of prior distributions \eqref{gout}--\eqref{gs2}. While $p_{x_{kt}}(x_{kt})$ can be obtained from the transmitters, parameters $\sigma^2$, $\varphiv$, $p_{01}$, and $\lambda$ in $p(\Yv|\Zv)$, $p(\Sv|\Cv)$, and $p(\Cv)$ are usually difficult to extract before the {estimation} procedure. Besides, as mentioned in Section \ref{signalmodel}, the AoA directions in $\varthetav$ may be difficult to acquire in general. Following the EM principle \cite{DL_Goodfellow}, we infer the posterior distribution $p(\Xv,\Sv,\Cv|\Yv)$ with unknown parameters $\Psi \triangleq \{\sigma^2,\lambda, p_{01}, \varphiv, \varthetav\}$ by maximizing the corresponding evidence lower bound (ELBO) $\mathcal{L}(\Yv,\Psi,q)$, i.e.
 \begin{align}
\label{elbo}
 p(\Xv,\Sv,\Cv|\Yv) = \argmax_{q} \underbrace{\E_{q}\left[\ln p(\Yv,\Xv,\Sv,\Cv;\Psi)\right]+H(q)}_{\mathcal{L}(\Yv,\Psi,q)},
\end{align}
where the expectation is w.r.t. an arbitrary probability distribution $q$ over $\Xv$, $\Sv$ and $\Cv$, and $H(\cdot)$ is the entropy function.

The EM algorithm maximizes $\mathcal{L}$ with respect to $q$ and $\Psi$ in an alternating fashion. The detailed steps are as follows.
\begin{itemize}
	\item E-step: Denote the initial guess of the parameters as $\Psi^{(0)}$. At the $j$-th iteration, set $q^{(j)}(\Xv,\Sv,\Cv|\Yv)=p(\Xv,\Sv,\Cv|\Yv;\Psi^{(j)})$. In other words, $q$ is defined as the posterior in terms of the current value of $\Psi$.
	\item M-step: Update $\Psi^{(j+1)}$ by
	\begin{align}
	\label{em1}
	\Psi^{(j+1)}&=\argmax_{\Psi}\mathcal{L}(\Yv,\Psi,q^{(j)})\nonumber\\
	&=\argmax_{\Psi}\E_{q^{(j)}}\left[\ln p(\Yv,\Xv,\Sv,\Cv;\Psi)\right].
	\end{align}
\end{itemize}
In the sequel, we discuss our design to realize the EM algorithm based on the idea of approximate inference.
\subsection{Approximate-Inference-Based EM Algorithm \label{EM}}
The posterior $q^{(j)}$ in each iteration is generally intractable even for given $\Psi^{(j)}$. We realize the E-step by restricting a particular family of distribution $q$, a.k.a. approximate inference\cite{DL_Goodfellow}. Here, we set $q$ as a factorizable distribution. In other words, aforementioned $q^{(j)}$ is approximated by the product of the marginal distributions as
 \begin{align}
\label{em2}
q^{(j)}&(\Xv,\Sv,\Cv|\Yv)\nonumber\\
&=\prod_{k=1}^K\prod_{t=1}^T p(x_{kt}|\Yv;\Psi^{(j)})\prod_{l=1}^L\prod_{k=1}^K p(s_{lk},\!c_{lk}|\Yv;\Psi^{(j)}).
\end{align}
 \remark{The proposed approximation in \eqref{em2} seems similar to, but is different from the idea of variational inference. Specifically, variational inference aims to find the factorizable distribution $\tilde{ q}(\Xv,\Sv,\Cv)\triangleq\prod_{k=1}^K\prod_{t=1}^T q_{kt}(x_{kt})\prod_{l=1}^L\prod_{k=1}^Kq_{lk}(s_{lk},c_{lk})$, such that the Kullback-Leibler divergence w.r.t. the target distribution $\text{KL}(\tilde{ q}(\Xv,\Sv,\Cv)\!\!\parallel \!\!p(\Xv,\Sv,\Cv|\Yv;\Psi^{(j)}))$ is minimized. However, here we approximate the true posterior as the product of marginal distributions. It is shown in \cite{PRML} that the proposed approximation corresponds to the minimizer of the \emph{inverse Kullback-Leibler divergence} $\text{KL}(p(\Xv,\Sv,\Cv|\Yv;\Psi^{(j)})\!\parallel \!q^{(j)}(\Xv,\Sv,\Cv|\Yv))$.}


We adopt the incremental variant scheme \cite{NN_EM2} and perform coordinate-wise maximization to update the elements in $\Psi$ sequentially. Specifically, we have the following update rules for the M-step:
\begin{subequations}
\begin{align}
(\sigma^2)^{(j+1)}\!=\!&\argmax_{\sigma^2}\E\left[\ln p(\Yv,\!\Xv,\!\Sv,\!\Cv;\sigma^2\!,\!\varthetav^{(j)}\!,\!\varphiv^{(j)}, p_{01}^{(j)}\!,\! \lambda^{(j)}\!) \right]\!;\label{up1}\\
\vartheta_l^{(j+1)}\!=\!&\argmax_{\vartheta_l}\E\Big[\ln p(\Yv,\!\Xv,\!\Sv,\!\Cv;(\sigma^2)^{(j\!+\!1)}\!,\!\vartheta_1^{(j\!+\!1)}\!,\cdots, \nonumber\\
&\vartheta_{l\!-\!1}^{(j\!+\!1)},\vartheta_l,\vartheta_{l\!+\!1}^{(j)},\cdots,\vartheta_L^{(j)}, \varphiv^{(j)},p_{01}^{(j)},\lambda^{(j)} ) \Big], \forall l;\label{up2}\\
\varphi_k^{(j+1)}\!=\!&\argmax_{\varphi_k}\E\Big[\ln p(\Yv,\!\Xv,\!\Sv,\!\Cv;\sigma^{(j\!+\!1)},\varthetav^{(j\!+\!1)},\varphi_1^{(j\!+\!1)},\nonumber\\
&\cdots,\varphi_{k\!-\!1}^{(j\!+\!1)},\varphi_k,\varphi_{k\!+\!1}^{(j)},\cdots,\varphi_K^{(j)},p_{01}^{(j)}, \lambda^{(j)}) \Big],\forall k;\label{up3}\\
p_{01}^{(j+1)}\!=\!&\argmax_{p_{01}}\E\Big[\ln p(\Yv,\!\Xv,\!\Sv,\!\Cv;(\sigma^2)^{(j\!+\!1)},\varthetav^{(j\!+\!1)},\nonumber\\
& \varphiv^{(j\!+\!1)}, p_{01},\lambda^{(j)})\Big];\label{up4}\\
\lambda^{(j+1)}=&\argmax_{\lambda}\E\Big[\ln p(\Yv,\!\Xv,\!\Sv,\!\Cv;(\sigma^2)^{(j\!+\!1)},\varthetav^{(j\!+\!1)},\nonumber\\ &\varphiv^{(j\!+\!1)}, p_{01}^{(j\!+\!1)},\lambda) \Big],\label{up5}
\end{align}
\end{subequations}
where the expectations are w.r.t. $q^{(j)}$ defined in \eqref{em2}. Note that once we have the exact form of $q^{(j)}$, \eqref{up1}--\eqref{up5} can be solved efficiently by fixed-point equations. For example, we can solve \eqref{up1}--\eqref{up5} by forcing the derivative of the objective to zero with respect to the corresponding variable.

The performance of coordinate ascent algorithms may be compromised when the optimized coordinates are correlated. As seen from the probabilistic model in Section \ref{mp}, $\varphiv$, $\lambda$, and $p_{01}$ are tightly coupled. To avoid an oscillation in the iterative process, we update $\varphiv$, $\lambda$, and $p_{01}$ with multiple times in each iteration to improve the convergence of the overall algorithm.  
\section{Computation of Marginals and Other Implementation Details\label{section4}}
There exist a number of techniques to calculate or approximate the marginal posteriors required in \eqref{em2}, such as calculus of variants, sparse Bayesian learning, and message passing. Here, we propose to compute the marginal posteriors in \eqref{em2} by message passing, which is used to execute the M-step in \eqref{up1}--\eqref{up5}.
\subsection{Computation of Marginal Posteriors\label{amp}}
\begin{figure}[!t]
\centering
\includegraphics[width=3.6in]{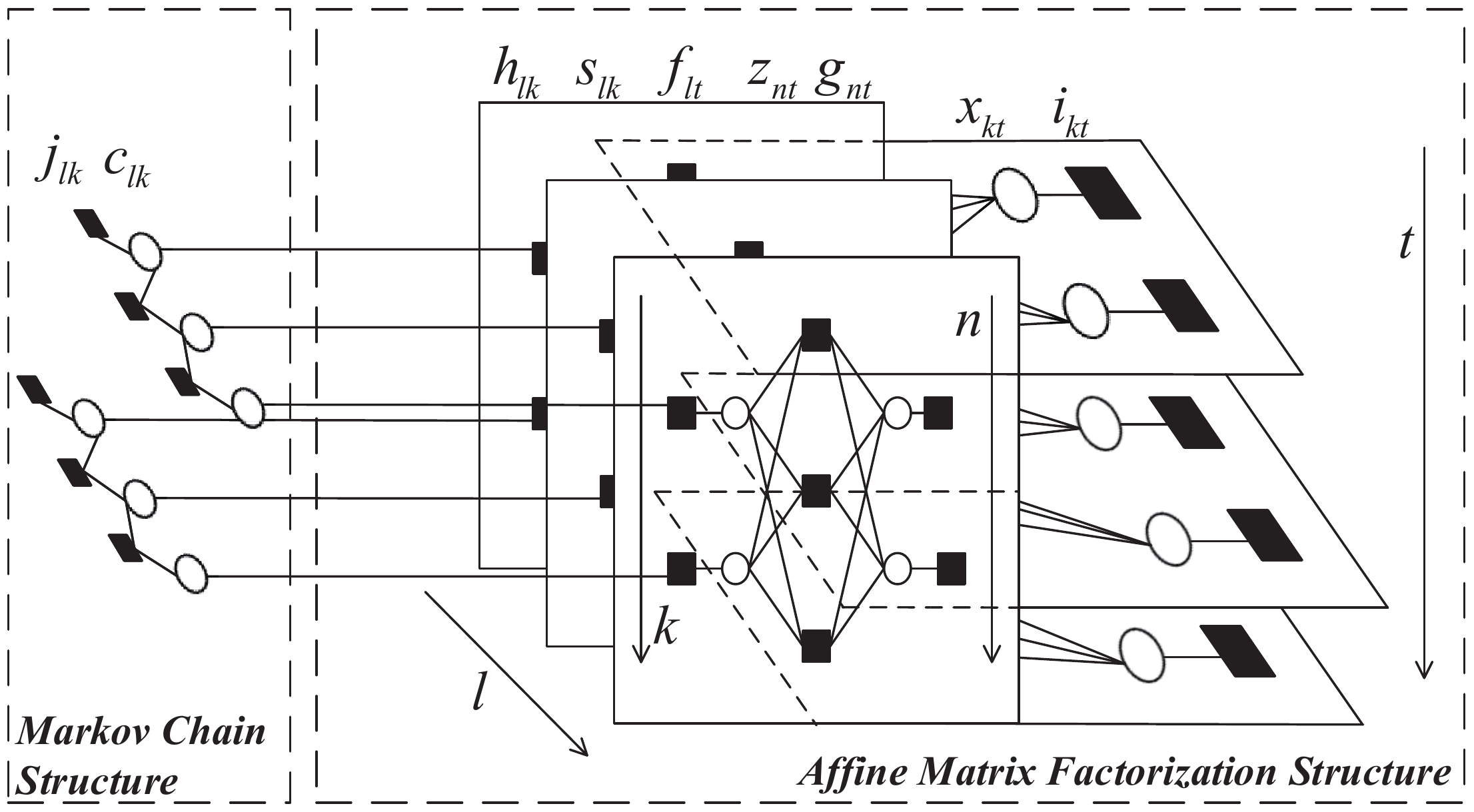}
\caption{An example of the factor graph representation for $T=L=3$ and $N=K=2$, where empty circles and filled squares represent variable nodes and factor nodes, respectively. }
\label{fg1}
\end{figure}

From the model introduced in Section \ref{mp}, we obtain
\begin{align}
p( \Yv,&\Xv,\Sv,\Cv)\propto p(\Yv|\Zv)p(\Xv)p(\Sv|\Cv)p(\Cv)\delta(\Zv-\Av(\varthetav)\Sv\Xv)\nonumber\\
\propto&\left( \prod_{n=1}^N\prod_{t=1}^T  p(y_{nt}|z_{nt})\delta\!\left(\! z_{nt}\!-\!\sum_{l=1}^L\sum_{k=1}^K A_{nl}s_{lk}x_{kt}\right) \right) \nonumber\\
&\left(\prod_{k=1}^K\prod_{t=1}^T  p_{x_{kt}}(x_{kt})\right)  \left(\prod_{l=1}^L\prod_{k=1}^K p(s_{lk}|c_{lk})p(c_{lk})\right).\label{bayes}
\end{align}
The factor graph representation of \eqref{bayes} is depicted in Fig. \ref{fg1}. The factorized pdfs, represented by factor nodes, are connected with their associated arguments, represented by variable nodes. We divide the whole factor graph into two sub-regions to distinguish the AMF structure for \eqref{Y1} and the Markov chain (MC) structure for \eqref{gc2}. We summarize the notations of the factors in Table \ref{BPnotations}. 

The message passing algorithm on the factor graph in Fig. \ref{fg1} is described as follows. Denote by $\Delta^i_{a \to b}(\cdot)$ the message from node $a$ to $b$ in iteration $i$, and by $\Delta_{c}^i(\cdot)$ the marginal posterior computed at variable node $c$ in iteration $i$. Applying the sum-product rules, we obtain the following messages and marginal posteriors.
\subsubsection{Messages within the AMF structure} For $\forall n,t,l,k,$
\begin{align}
\Delta_{z_{nt}\to f_{lt}}^i (w_{lt})\propto& \int_{\{w_{l^\prime \! t}\}_{l^\prime \! \neq l}}\!p_{y_{nt}|z_{nt}}(y_{nt}|\sum_{l=1}^L A_{nl}w_{lt})\nonumber \\
&\prod_{l^\prime \! \neq l} \Delta^i_{f_{l^\prime \! t}\to z_{nt}}(w_{l^\prime \! t});   \label{ztow}\\
\Delta_{ f_{lt}\to z_{nt}}^{i\!+\!1}(w_{lt})\propto& p^i_{w_{lt}}(w_{lt}) \prod_{n^\prime \! \neq n}\Delta^i_{z_{n^\prime \! t}\to f_{lt}}(w_{lt}); \label{wtoz}\\
\Delta_{ f_{lt}\to x_{kt}}^i(x_{kt})\propto & \int_{\{s_{lk}\}_{\forall k},\{x_{k^\prime \! t}\}_{k^{\prime} \! \neq  k}} \prod_{k=1}^K \Delta_{s_{lk}\to f_{lt}}^i(s_{lk})\nonumber\\
& \prod_{k^\prime \!\neq  k}\Delta_{x_{k^\prime \! t}\to f_{lt}}^i(x_{k^\prime \! t})\prod_{n=1}^N\Delta_{z_{nt}\to f_{lt}}^i (w_{lt}); \label{wtox}
\end{align}
\begin{align}
\Delta_{ x_{kt} \to  f_{lt}}^{i\!+\!1}(x_{kt})\propto & p_{x_{kt}}(x_{kt}) \prod_{l^\prime \! \neq  l}\Delta_{ f_{l^\prime \! t}\to x_{kt}}^i(x_{kt});\label{xtow}\\
\Delta_{ f_{lt}\to s_{lk}}^i(s_{lk})\propto & \int_{\{x_{kt}\}_{\forall k},\{s_{lk^\prime}\!\}_{k^{\prime} \! \neq  k}} \prod_{k=1}^K \Delta_{x_{kt}\to f_{lt}}^i(x_{kt})\nonumber\\
& \prod_{k^\prime \!\neq  k}\Delta_{s_{lk^\prime}\!\to f_{lt}}^i(s_{lk^\prime})\prod_{n=1}^N\Delta_{z_{nt}\to f_{lt}}^i (w_{lt}); \label{wtos}\\
\Delta_{ s_{lk} \to  f_{lt}}^{i\!+\!1}(s_{lk})\propto & \Delta_{h_{lk} \to s_{lk}}^i(s_{lk}) \prod_{t^\prime \! \neq  t}\Delta_{ f_{l t^\prime}\to s_{lk}}^i(s_{lk}),\label{stow}
\end{align}
where $p^i_{w_{lt}}(w_{lt})$ in \eqref{wtoz} is defined to be 
\begin{align}
p^i_{w_{lt}}(w_{lt})\propto&\int_{\{s_{lk},x_{kt}\}_{\forall k}} \! \left(\prod_{k=1}^K  \Delta^i_{x_{kt}\to f_{lt}}(x_{kt})\Delta^i_{s_{lk}\to f_{lt}}(s_{lk})\right) \nonumber \\
&\delta\left( w_{lt}\!-\!\sum_{k=1}^Ks_{lk}x_{kt}\right) .\label{pw}
\end{align}
\begin{table}[!t]
	\centering
	\caption{Notation for the factor nodes}\label{BPnotations}
	\begin{tabular}{l|l|l}
		\toprule
		Factor& Distribution & Exact Form\\
		\midrule
		$f_{lt}$&$p(w_{lt}|s_{lk},x_{kt},\!\forall k)$&$\delta(w_{lt}-\sum_{k}s_{lk}x_{kt})$\\
		$g_{nt}$&$p(y_{nt}|z_{nt})$&$\CN(y_{nt};z_{nt},\sigma^2)$\\
		$h_{lk}$&$p(s_{lk}|c_{lk})$&$\delta\left( c_{lk}\right) \delta\left( s_{lk}\right)\!+$\\
		&&$\delta\left(c_{lk}\!-\!1\right)\CN(s_{lk};0,\varphi_k)$\\
		$i_{kt}$&$p_{x_{kt}}\left( x_{kt}\right)$&$p_{x_{kt}}\left( x_{kt}\right)$\\
		$j_{lk}$&$p\left( c_{lk}|c_{l\!-\!1,k}\right)$&$\begin{cases}
		(1\!-\!p_{10})^{1\!-\!c_{lk}}p_{10}^{c_{lk}},c_{l-1,k}\!=\!0\\
		p_{01}^{1\!-\!c_{lk}}(1\!-\!p_{01})^{c_{lk}},c_{l-1,k}\!=\!1
		\end{cases}$\\
		\bottomrule
	\end{tabular}
\end{table}

\subsubsection{Messages within the MC structure} For $\forall l,k,$
for $l=1,\cdots,L-1$,
\begin{align}
&\Delta_{j_{l\!+\!1,k}\to c_{l\!+\!1,k}}^i(c_{l\!+\!1,k})\nonumber\\
&\propto \!\sum_{c_{lk}\in\{0,1\}}p\left( c_{l\!+\!1,k}|c_{lk}\right)\Delta_{j_{lk}\!\to\! c_{lk}}^i(c_{lk})\Delta_{h_{lk}\!\to\!c_{lk} }^i(c_{lk}),\label{jtoc}
\end{align}
and $\Delta_{j_{1k}\!\to\! c_{1k}}^i(c_{1k})$ is set to $(1\!-\!\lambda)\delta(c_{1k})\!+\!\lambda\delta(c_{1k}\!-\!1)$.

For $l=L,L-1,\cdots,2$,
\begin{align}
&\Delta_{j_{lk}\to c_{l\!-\!1,k}}^i(c_{l\!-\!1,k})\nonumber\\
&\propto \!\sum_{c_{lk}\in\{0,1\}}p\left( c_{lk}|c_{l\!-\!1,k}\right)\Delta_{j_{l\!+\!1,k}\!\to\! c_{lk}}^i(c_{lk})\Delta_{h_{lk}\!\to\!c_{lk} }^i(c_{lk}),\label{ctoj}
\end{align}
and $\Delta_{j_{L\!+\!1,k}\!\to\! c_{Lk}}^i(c_{Lk})$ is set to $\frac{1}{2}\delta(c_{Lk})\!+\!\frac{1}{2}\delta(c_{Lk}\!-\!1)$.
\begin{table}[!t]
	\centering
	\caption{Notation of mean and variance for various messages and posteriors}\label{table1.5}
	\begin{tabular}{l|l|l}
		\toprule
		Message/Posterior& Mean & Variance\\
		\midrule
		$\Delta_{f_{lt}\to z_{nt}}^i(w_{lt})$&$\wh_{lt,n}(i)$&$v^w_{lt,n}(i)$\\
		$\Delta_{x_{kt}\to f_{lt}}^i(x_{kt})$&$\xh_{kt,l}(i)$&$v^x_{kt,l}(i)$\\
		$\Delta_{s_{lk}\to f_{lt}}^i(s_{lk})$&$\sh_{lk,t}(i)$&$v^s_{lk,t}(i)$\\
		$\Delta_{x_{kt}}^i(x_{kt})$&$\xh_{kt}(i)$&$v^x_{kt}(i)$\\
		$\Delta_{s_{lk}}^i(s_{lk})$&$\sh_{lk}(i)$&$v^s_{lk}(i)$\\
		\bottomrule
	\end{tabular}
\end{table}
\begin{table}[!t]
	\centering
	\caption{Messages and posteriors after approximations}\label{table2}
	\begin{tabular}{l|l}
		\toprule
		Message/Posterior& Functional Form\\
		\midrule
		$\Delta_{h_{lk}\to c_{lk}}^i(c_{lk})$&$\pi^{\text{out}}_{lk}(i)$\\
		$\Delta_{c_{lk}\to h_{lk}}^i(c_{lk})$&$\pi^{\text{in}}_{lk}(i)$\\
		$\Delta_{j_{lk}\to c_{lk}}^i(c_{lk})$&$\lambda^{\text{fwd}}_{lk}(i)$\\
		$\Delta_{j_{l+1,k}\to c_{lk}}^i(c_{lk})$&$\lambda^{\text{bwd}}_{lk}(i)$\\
		$\Delta_{ c_{lk}}^i(c_{lk})$&$\omega_{lk}(i)$\\
		$\Delta_{s_{lk}\to h_{lk}}^i(s_{lk})$&$\CN(s_{lk};\qh_{lk}(i),v^q_{lk}(i))$\\
		$\Delta_{h_{lk}\to s_{lk}}^i(s_{lk})$&$\pi^{\text{in}}_{lk}(i)\CN(s_{lk};0,\varphi_k)\!+\!$\\
		&$(1\!-\!\pi^{\text{in}}_{lk}(i))\delta(s_{lk})$\\
		$\Delta_{f_{lt}\to z_{nt}}^i(w_{lt})$&$\CN(w_{lt};\wh_{lt,n}(i),v^w_{lt,n}(i))$\\
		$\Delta_{x_{kt}\to f_{lt}}^i(x_{kt})$&$\frac{1}{Z}\CN(x_{kt};\rh_{kt,l}(i),\!v^r_{kt,l}(i))p_{x_{kt}}\!\left( x_{kt}\right)$\\
		$\Delta_{s_{lk}\to f_{lt}}^i(s_{lk})$&$\frac{1}{Z}\CN(s_{lk};\qh_{lk,t}(i),\!v^q_{lk,t}(i))\Delta_{h_{lk}\!\to\! s_{lk}}^i\!(s_{lk})$\\
		$\Delta_{x_{kt}}^i(x_{kt})$&$\frac{1}{Z}\CN(x_{kt};\rh_{kt}(i),\!v^r_{kt}(i))p_{x_{kt}}\!\left( x_{kt}\right)$\\
		$\Delta_{s_{lk}}^i(s_{lk})$&$\frac{1}{Z}\CN(s_{lk};\qh_{lk}(i),\!v^q_{lk}(i))\Delta_{h_{lk}\!\to\! s_{lk}}^i\!(s_{lk})$\\
		\bottomrule
	\end{tabular}
\end{table}
\subsubsection{Messages exchanged between the MC structure and the AMF structure} For $\forall l,k,$
\begin{align}
\Delta_{c_{lk}\to h_{lk}}^{i}(c_{lk})\propto & \Delta_{j_{lk}\to c_{lk}}^i(c_{lk})\Delta_{j_{l\!+\!1, k}\to c_{lk}}^i(c_{lk});\label{ctoh}\\
\Delta_{h_{lk}\to c_{lk}}^{i\!+\!1}(c_{lk})\propto & \int_{s_{lk}}p(s_{lk}|c_{lk})\prod_{t=1}^T \Delta_{ f_{lt}\to s_{lk}}^i(s_{lk}).\label{htoc}
\end{align}
\subsubsection{Marginal functions at variable nodes} For $\forall l,k,t,$
\begin{align}
\Delta_{ x_{kt}}^{i\!+\!1}(x_{kt})\propto & p_{x_{kt}}(x_{kt}) \prod_{l=1}^L \Delta_{ f_{lt}\to x_{kt}}^i(x_{kt});\label{xmarg}\\
\Delta_{ s_{lk}}^{i\!+\!1}(s_{lk})\propto & \Delta_{h_{lk} \to s_{lk}}^i(s_{lk}) \prod_{t=1}^T \Delta_{ f_{l t}\to s_{lk}}^i(s_{lk});\label{smarg}\\
\Delta_{c_{lk}}^{i\!+\!1}(c_{lk})\propto & \Delta_{h_{lk}\to c_{lk}}^{i}(c_{lk}) \Delta_{j_{lk}\to c_{lk}}^i(c_{lk})\Delta_{j_{l\!+\!1, k}\to c_{lk}}^i(c_{lk}).\label{cmarg}
\end{align}

To reduce the overall computation complexity, we employ the ``turbo" scheme \cite{MIMO_Turbo} to schedule message updating.
For the $j$-th EM iteration, we first perform message passing within the AMF structure until the updating process converges or reaches the maximum number of loops. Then, the updated messages are passed into the MC structure, where the updates within the MC structure are executed.
\begin{algorithm}[!t]
	\caption{Message Passing for the E-Step}
	\label{alg1}
	\textbf{Input:} $\Yv; \Av(\varthetav); \sigma^2; p_{x_{kt}}\left( x_{kt}\right); \varphiv; \lambda; p_{01}.$\\
	\textbf{Initialization}:
	$\betah_{nt}(0)=\gammah_{lt}(0)=0$;\\
	$v^w_{lt}(1)=v^s_{lk}(1)=v^x_{kt}(1)=10$; $\wh_{lt}(1)=\sh_{lk}(1)=0$;\\
	$\xh_{kt}(1)$ randomly drawn from $p_{x_{kt}}\left( x_{kt}\right)$.
	\begin{algorithmic}[0]
		\FOR{$i=1,2,\cdots,I_{\text{max}}$}
		\STATE
		{
			\%Update messages within the AMF structure:\\
			For $\forall l,t$, update $\wh_{lt}(i\!+\!1)$ and $v^w_{lt}(i\!+\!1)$ by \eqref{vuh_nn}--\eqref{zetah}, \eqref{vph2}--\eqref{ph2}, and \eqref{vwh_nn}--\eqref{wh_nn};\\
			\%Update the posteriors:\\
			For $\forall l,k,t$, update $\Delta_{x_{kt}}^{i\!+\!1}(x_{kt})$ and $\Delta_{s_{lk}}^{i\!+\!1}(s_{lk})$ as in Table \ref{table2} by \eqref{valphah2}--\eqref{qh2}, then update the means and variances by \eqref{xposta}--\eqref{sposta};\\
			\IF{$\sqrt{\frac{\sum_{k}\sum_{t}\abs{\xh_{kt}(i\!+\!1)-\xh_{kt}(i)}^2 }{\sum_{k}\sum_{t}\abs{\xh_{kt}(i)}^2 }}\leq \delta_{1}$} 
			\STATE{stop;}   \ENDIF     
		} 
		\ENDFOR\\
		\%Update $\Delta_{h_{lk}\to c_{lk}}^i$ and $\Delta_{c_{lk}\to h_{lk}}^i$:\\
		For $\forall l,k$, update $\pi^{\text{out}}_{lk}(i)$ and $\pi^{\text{in}}_{lk}(i)$ by \eqref{piout}--\eqref{piin};\\
		\%Update $\Delta_{j_{lk}\to c_{lk}}^i$ and $\Delta_{j_{l+1,k}\to c_{lk}}^i$:\\ 
		For $\forall l,k$, update $\lambda^{\text{fwd}}_{lk}(i)$ and $\lambda^{\text{bwd}}_{lk}(i)$ by \eqref{fwd1}--\eqref{bwd};\\
		\%Update $\Delta_{ c_{lk}}^{i\!+\!1}$:\\
		For $\forall l,k$, update $\omega_{lk}(i\!+\!1)$ by \eqref{temp64}.\\
	\end{algorithmic}
	\textbf{Output:}{$\Delta^{i\!+\!1}_{x_{kt}}$ and $\Delta^{i\!+\!1}_{s_{lk}}$ as the estimations of $p(x_{kt}|\Yv; \Psi)$ and $p(s_{lk}|\Yv,\Psi)$, respectively.
}
\end{algorithm}

Message passing within the AMF structure requires high-dimensional integration and normalization. To reduce the computation complexity, we approximately calculate \eqref{ztow}--\eqref{stow}  in the large-system limit, \ie $N,K,L,T \to \infty$ with fixed ratios for $L/N$, $K/N$ and $T/N$, following the general idea of the AMP framework \cite{MP_AMP1,MP_GAMP}. Without loss of generality, we assume that $x_{kt}$ scales as $O(1/\sqrt{T})$, $A_{nl}$ scales as $O(1/\sqrt{N})$, and other quantities scale as $O(1)$, which is the same as in \cite{MP_AMP1,MP_GAMP, MP_BIGAMP1}.

 To facilitate the approximations, we define the mean and variance quantities in Table \ref{table1.5}.  We sketch the major approximations in the following, where the rigorous derivations can be found in Appendix \ref{mpproof}.

\begin{itemize}
	\item A second-order Taylor expansion together with the Gaussian integral is employed to derive the tractable closed-form approximation for \eqref{pw};
	\item Based on the central limit theorem \cite{MP_AMP1}, we impose the Gaussian approximation on the product of a large number of messages, such as $\prod_{l^\prime \! \neq l} \Delta^i_{f_{l^\prime \! t}\to z_{nt}}$ in \eqref{ztow}. Similar arguments also apply to the terms in \eqref{wtox} and \eqref{wtos};
	\item Second-order Taylor expansions are introduced in the computations of \eqref{ztow}, \eqref{wtox}, and \eqref{wtos} to further simplify the resulting messages;
	\item To close the loop, we neglect some vanishing terms and obtain the ``Onsager term" as in \cite{MP_AMP1,MP_GAMP}; see, \eg \eqref{uh_nn} and \eqref{ph2}.
\end{itemize}

The resultant messages and posteriors after the approximations are shown in Table \ref{table2}, where $Z$ is the normalization factor ensuring the messages are integrated to $1$. For brevity, we only show the nonzero probability for Bernoulli distributions, \eg $\pi^{\text{out}}_{lk}(i)\triangleq \Delta_{h_{lk}\to c_{lk}}^i(c_{lk}=1)$. In Algorithm \ref{alg1}, we summarize the detailed computations of distributions listed in Table \ref{table2}. We add a threshold  $\delta_{1}$ for early stopping and use adaptive damping \cite{MP_MP0} to accelerate convergence.
\remark{We note that in \cite{MP_ARM}, the authors developed a message passing algorithm, termed GARM-AMP, to solve the affine rank minimization problem, which has a similar structure as the AMF problem in \eqref{Y1}. Although the algorithm in \cite{MP_ARM} appears similar to the approximated message passing formulas inside the AMF structure, there are substantial differences between our proposed algorithm and GARM-AMP: 
	\begin{itemize}
		\item In our problem, we have an additional MC structure, and hence a different factor graph and a different message passing scheme compared to \cite{MP_ARM}.
		\item We approximate particular messages inside the AMF structure based on the central limit theorem. For example, we argue that  $\prod_{l^\prime \! \neq l} \Delta^i_{f_{l^\prime \! t}\to z_{nt}}$ in \eqref{ztow} converges to the normal distribution in the large-system limit. Differently, GARM-AMP approximates messages by taking Taylor expansion on separate messages and sum them up with re-exponentiation; see, \eg eqs. (82)--(84) in \cite{MP_ARM}. 
		\item GRAM-AMP has two additional terms in \eqref{vrh2} and \eqref{vqh2}, whereas we take further approximations to eliminate those terms following eq. (121) in \cite{MP_BIGAMP1} and eqs. (113)--(114) in \cite{MP_BIGAMP3}. As pointed out in \cite{MP_BIGAMP3}, these approximations have a serious impact on the fixed point of the message passing algorithm. Moreover, the message updating orders are different. For example, Algorithm \ref{alg1} updates \eqref{vwh_nn}--\eqref{wh_nn} before the execution of \eqref{valphah2}--\eqref{qh2}, while GARM-AMP updates the terms corresponding to \eqref{vwh_nn}--\eqref{wh_nn} (\ie eqs. (A13)--(A14) in \cite{MP_ARM}) after finishing the computations for all intermediate mean and variance quantities. Finally, the damping schemes are different as well. GRAM-AMP only damps the equations corresponding to  \eqref{vwh_nn}--\eqref{wh_nn} and \eqref{xposta}--\eqref{sposta}, whereas we additionally damp \eqref{vuh_nn}, \eqref{vtauh}--\eqref{tauh}, \eqref{vph2}, and \eqref{valphah2}---\eqref{alphah2}.\footnote{Without these changes, GARM-AMP does not converge in the settings considered in this work. We guess that there could be some additional tricks in GARM-AMP that are not revealed by the authors in \cite{MP_ARM}.
}
	\end{itemize}
}
\subsection{Angle Tuning for Super-Resolution}
Equipped with the posteriors computed in Section \ref{amp}, we are ready to provide the solutions of \eqref{up1}--\eqref{up5} in the M-step. 
We name \eqref{up2} the angle-tuning procedure, which aims to increase the resolution of AoAs, \ie to reduce the mismatch between $\varthetav$ and true AoAs. \eqref{up2} is non-convex and the fixed-point solution is intractable in general. We develop a gradient ascend method to approximately solve \eqref{up2}. Specifically, we compute
\begin{equation}
\label{thetaupdate}
\vartheta^{(j+1)}_l=\vartheta^{(j)}_l+\epsilon \frac{\varpi_{MP}^{(j)}(\vartheta^{(j)}_l)}{\abs{\varpi^{(j)}_{MP}(\vartheta^{(j)}_l)} },
\end{equation}
where $\varpi_{MP}^{(j)}(\vartheta_l^{(j)})$ denotes the derivative of the objective in \eqref{up2} w.r.t. $\vartheta_l^{(j)}$, and $\epsilon$ is the step size. We show in Appendix \ref{pro4} that the gradient is given by \eqref{temp13}, shown on top of this page, 
\begin{figure*}[ht]
\begin{equation}
\begin{aligned}
\label{temp13}
\varpi_{MP}^{(j)}(\vartheta_l^{(j)})\!=&2\left( \sigma^{(j)}\right)^{-2}\Re\left(((\av^{(j)}_l)^\prime)^H \left(\sum_{t=1}^T\left( \wh_{lt}^{(j)}\right)^\star \yv^{(j)}_{t-l}\right)\!-\!\left(((\av^{(j)}_l)^\prime)^H\av_l^{(j)}\right)\left( \sum_{t=1}^T \left(\left( v^w_{lt}\right) ^{(j)}\!+\!\abs{ \wh_{lt}^{(j)}}^2\right)\right)\right),
\end{aligned}
	\end{equation} 
\hrulefill
\end{figure*}
where $\Re(\cdot)$ represents the real part of a complex number; $\av_l^{(j)}$ denotes the $l$-th column of $\Av(\varthetav^{(j)})$; $(\av^{(j)}_l)^\prime=\frac{\partial \av_l}{\partial \vartheta_l}\big|_{\vartheta_l=\vartheta_l^{(j)}}$; and  $\yv_{t-l}^{(j)}=\yv_t\!-\!\sum_{l^\prime \neq l}\left( \wh_{l^\prime t}\right) ^{(j)}\av_{l^\prime}^{(j)}$.

Gradient ascent is known to have a relatively slow convergence rate. Additional adjustments are needed to guarantee the convergence in the practical implementation. Firstly, we employ a fixed step size $\epsilon\!=\!\frac{\pi}{2tL}$ for some integer $t$. Secondly, we update $\varthetav$ twice according to \eqref{thetaupdate} in each EM iteration. The choice of $\epsilon$ and the nested update  guarantee that the true AoAs can be approached within $t$ iterations.

\subsection{Learning Other Parameters}
The remaining optimization problems can be solved by fixed-point equations. We show in Appendix \ref{pro5} that the solutions to \eqref{up1}, \eqref{up3}--\eqref{up5} are given by
\begin{subequations}
	\begin{align}
	(\sigma^2)^{(j+1)}\!&=\!\frac{1}{NT}\left(\norm{\Yv-\Zvh^{(j)}}_F^2+\sum_{n=1}^N\sum_{t=1}^T (v^z_{nt})^{(j)}\right),\label{sigmaupdate}\\
	\varphi_k^{(j+1)}\!&=\!\frac{\sum_{l=1}^L\eta_{lk}^{(j)}\left( \abs{ \chi_{lk}^{(j)}}^2\!+\! \nu_{lk}^{(j)}\right) }{\sum_{l=1}^L\eta_{lk}^{(j)}},\label{varphiupdate}\\
	p_{01}^{(j+1)}\!&=\!1-\!\frac{\sum_{l=1}^{L-1}\sum_{k=1}^K p(c_{l\!+\!1,k}\!=\!1,c_{lk}\!=\!1|\Yv,\Psi^{(j)})}{\sum_{l=1}^{L-1}\sum_{k=1}^K\omega_{lk}^{(j)}},\label{p01update}\\
	\lambda^{(j+1)}\!&=\!\frac{1}{K}\sum_{k=1}^K \omega_{1k}^{(j)},\label{lambdaupdate}
	\end{align}
\end{subequations}
where
\begin{subequations}
\begin{align}
\eta_{lk}^{(j)}&=\frac{(\pi^{\text{out}}_{lk})^{(j)}(\pi^{\text{in}}_{lk})^{(j)}}{(\pi^{\text{out}}_{lk})^{(j)}(\pi^{\text{in}}_{lk})^{(j)}\!+\!(1\!-\!(\pi^{\text{out}}_{lk})^{(j)})(1\!-\!(\pi^{\text{in}}_{lk})^{(j)})},\label{temp55}\\
\nu_{lk}^{(j)}&=\varphi_{k}^{(j)}(v^q_{lk})^{(j)}/\left(\varphi_{k}^{(j)}\!+\!(v^q_{lk})^{(j)} \right) ,\label{temp56}\\
 \chi_{lk}^{(j)}&=\varphi_{k}^{(j)}\qh_{lk}^{(j)}/\left(\varphi_{k}^{(j)}\!+\!(v^q_{lk})^{(j)} \right).\label{temp57}
\end{align}
\end{subequations}
\subsection{The Overall Algorithm} 
\label{ambig}
There exist ambiguities in the solution of problem \eqref{Y1}.
Specifically, if ($\Svh$, $\Xvh$) is a solution to \eqref{Y1}, so is ($\Svh \Uv$, $\Uv^{-1} \Xvh$), provided that $\Uv$ is a scaled permutation matrix, \ie $\Uv=\Xiv\Qv$ with a phase-shift diagonal matrix $\Xiv$ and a permutation matrix $\Qv$. 
\begin{algorithm}[!t]
	\caption{The AEM-MP Algorithm}
	\label{al3}
	\textbf{Input:} {$\Yv$; $\Av(\varthetav)$; and $x_{ref}$.}\\
	\textbf{Initialization}:$\lambda^{(0)}\!=\!0.1$; $p_{01}^{(0)}\!=\!0.5$; $(\sigma^2)^{(0)}\!=\!\frac{\norm{\Yv}_F^2}{100NT}$; $\varphiv^{(0)}\!=\!\bf{1}$; and $\varthetav^{(0)}$ uniformly samples the AoA range.\\
	Perform the projection by $\Vv_1$; \\
	\begin{algorithmic}[0]
		\FOR{$j=0,1,2,\cdots,J_{\text{max}}$}
		\STATE
		{
			Compute posteriors via message passing as in Algorithm \ref{alg1} and execute the E-Step using \eqref{em2};\\
			Execute the M-step by \eqref{thetaupdate} and \eqref{sigmaupdate}--\eqref{lambdaupdate};\\
			Estimate $\Xvh$ and $\Svh$ by \eqref{map} and perform the reverse projection by $\Vv_1^H$;\\
			Eliminate ambiguities by \eqref{ambi}--\eqref{ambi2};\\
			\IF{$\sqrt{\frac{\sum_{k}\sum_{t}\abs{\xh_{kt}(j\!+\!1)-\xh_{kt}(j)}^2 }{\sum_{k}\sum_{t}\abs{\xh_{kt}(j)}^2 }}\leq \delta_2$}
			\STATE{stop;}  \ENDIF     
		} 
		\ENDFOR
	\end{algorithmic}
	\textbf{Output:}{$\Xvh$ and $\Svh$.}
\end{algorithm}

Similar ambiguity problems have been studied previously in the literature \cite{MIMO_EVD} and \cite{MIMO_zhangjianwen}. Following \cite{MIMO_zhangjianwen}, we resolve the permutation ambiguity by inserting an identification label into the transmitted signals. Meanwhile, the phase ambiguity can be resolved by using a reference symbol \cite{MIMO_EVD}. Without loss of generality, we assume that the first transmitted symbol for each user is a reference symbol, \ie $x_k(1)=x_{\text{ref}}, \forall k \in [K]$. The phase ambiguity can be eliminated by computing the relative phase shift as
\begin{subequations}
\begin{align}
\xh_{kt}^\prime&= \frac{\xh_{kt}x_\text{ref}}{\xh_{k1}},\label{ambi}\\
\sh_{lk}^\prime&=\frac{\sh_{lk}\xh_{k1}}{x_{\text{ref}}}.\label{ambi2}
\end{align}
\end{subequations}
Besides, a massive MIMO system usually satisfies $T > K$, where the direct factorization of $\Yv$ may fail. Therefore, we follow the approach in \cite{MIMO_zhangjianwen} to project $\Yv \in \Complex^{N\times T}$ and $\Xv \in \Complex^{K\times T}$ into subspaces $\bar \Yv \in \Complex^{N\times K}$ and $\bar \Xv \in \Complex^{K\times K}$, respectively. We achieve this goal by the right multiplication with a semi-unitary matrix $\Vv_1 \in \Complex^{T\times K}$, which consists of $K$ right-singular vectors of $\Yv$ corresponding to the $K$ largest singular values of $\Yv$.

We are now ready to summarize the overall super-resolution blind channel-and-signal estimation algorithm in Algorithm \ref{al3}, namely the approximate-inference-based EM with message passing (AEM-MP). Note that an early stopping criteria can be added to Algorithm \ref{al3} to reduce the running time.
\section{Numerical Results}
\label{Simulation}
We conduct simulations to investigate the behavior of the proposed algorithm in this section.
The following state-of-the-art DFT-based algorithms are taken as baselines, all of which estimate the channel matrix and user signals under the virtual channel representation. Note that we apply the ambiguity elimination scheme described in Section \ref{ambig} to all approaches.
\begin{itemize}
	\item BiG-AMP\cite{MP_BIGAMP1}: $\Sv$ and $\Xv$ in \eqref{DFTeq} are recovered by BiG-AMP as the dictionary learning problem.
	\item Pro-BiG-AMP\cite{MIMO_zhangjianwen}: In contrast to BiG-AMP, additional projection for $\Yv$ is introduced before factorizing $\Sv$ and $\Xv$.
	\item Subspace-based method with channel sparsity (Sub-CS)\cite{MIMO_Amine}: Subspace projection scheme from \cite{MIMO_Neumann} is employed, except that additional $\ell_{1}$ regularization is added to force the sparsity of $\Sv$.
\end{itemize}
We define the evaluation metrics as the normalized mean square error (NMSE) of $\Xv$ and $\Hv$:
\begin{align}
\text{NMSE of }{\Xv}=\frac{\norm{\Xvh-\Xv}^2_F}{\norm{\Xv}^2_F},
\end{align}
\begin{align}
\text{NMSE of }{\Hv}=\frac{\norm{\Hvh-\Hv}^2_F}{\norm{\Hv}^2_F},
\end{align}
 where $\Hv\triangleq[\hvt_1, \hvt_2, \cdots, \hvt_K] \in \Complex^{N \times K}$ is the collection of the physical channel coefficients defined in \eqref{hktilde}.\footnote{Recall from \eqref{hktilde2} that we have $\Hv=\Av(\varthetav) \Sv$ for some angular response $\Av(\varthetav)$ and a sparse matrix $\Sv$. Note that $\Hv$ is not necessary to be sparse.}

We also compare the proposed algorithm with the following non-DFT based channel estimation algorithm.
\begin{itemize}
	\item Sparse Bayesian learning (SBL) based channel estimation  \cite{MIMO_DOA1}: Before the transmission of user signals, randomly generated training symbols with length $T_{\text{train}}$ are sent to the BS. The SBL-based channel estimation scheme is employed to estimate the channel matrix $\Hv$ prior to data detection.\footnote{The optimal training length $T_{\text{train}}$ for the SBL-based scheme is difficult to determine analytically. From numerical experiments, we find that $T_{\text{train}}\approx K$ is optimal in terms of the achievable rate in most cases. Therefore, we set $T_{\text{train}}= K$ in the implementation of the algorithm.}
\end{itemize}
\begin{figure}[!t]
	\centering
	\includegraphics[width=3.4in]{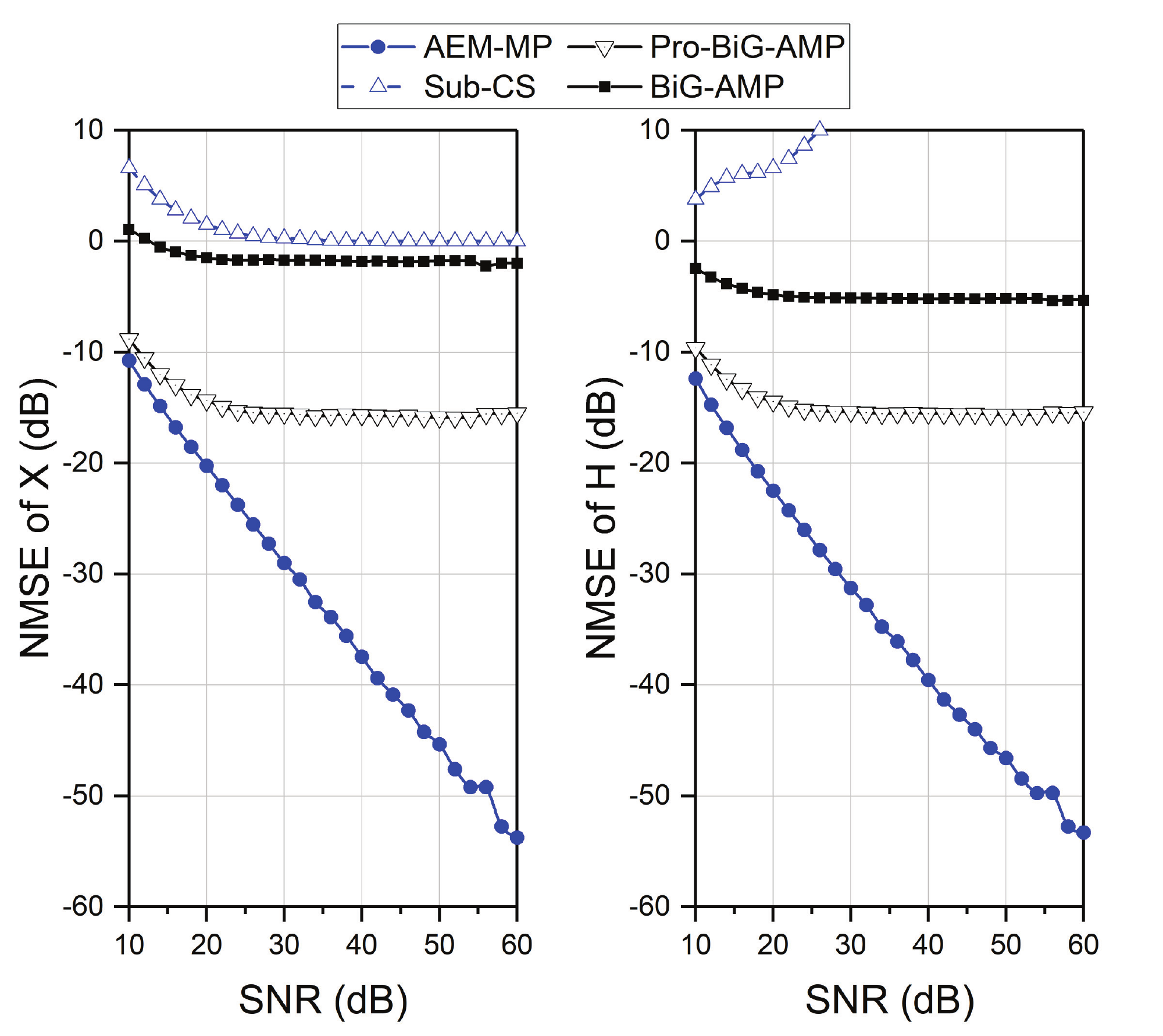}
	\caption{NMSE versus SNR with a fixed known grid, where $N\!=\!L\!=\!128$, $T\!=\!100$, $K\!=\!8$, and $\lambda\!=\!0.2$.}
	\label{snr_off}
\end{figure}

Note that the SBL-based scheme relies on the training symbols to estimate the channel. To facilitate the comparisons, we define the following achievable rates. For the blind channel-and-signal estimation system, the achievable rate is given by \cite{MIMO_zhangjianwen}
\begin{align}
\label{rate1}
R_{\text{blind}}=&\sum_{k=1}^K\left(1-\frac{1}{T}\right)  \log_2\left(1+\frac{\norm{\xv_k}_2^2}{\norm{\xv_k-\xvh_k}_2^2} \right)-\frac{K\lceil \log_2K\rceil}{T},
\end{align}
where the term $\left(1-1/T\right)$ is due to the single pilot symbol $x_{\text{ref}}$ in the phase ambiguity elimination, and the rate loss $\frac{K\lceil \log_2K\rceil}{T}$ is caused by the permutation ambiguity.
Likewise, the achievable rate for the training-based system is given by 
\begin{align}
\label{rate2}
R_{\text{training}}=&\sum_{k=1}^K\left(1-\frac{T_{\text{train}}}{T}\right)  \log_2\left(1+\frac{\norm{\xv_k}_2^2}{\norm{\xv_k-\xvh_k}_2^2} \right).
\end{align}

In the simulations, the user signals are \iid generated from the standard complex Gaussian distribution. Noise power is computed by a certain signal-to-noise ratio (SNR) level as $\sigma^2=K/\text{SNR}$. Channel vectors are generated based on \eqref{hktilde} with $\a_k(i,j)$ drawn from the standard complex Gaussian distribution. $L_c(k)$ and $L_p(k)$ are assumed to be the same for all users, and hence we drop the index $k$. We simulate ULAs and non-ULAs with different forms of $\av(\theta_{k}(i,j))$ in the sequel. For the AEM-MP algorithm in Algorithm \ref{al3}, $I_{\text{max}}$ is set to $300$ and $J_{\text{max}}$ is set to $14$. All the results are conducted by averaging over 1000 Monte Carlo trials, unless otherwise specified.
\subsection{Blind Channel-and-Signal Estimation with ULA}
\label{SIMU1}
\begin{figure}[!t]
	\centering
	\includegraphics[width=3.4in]{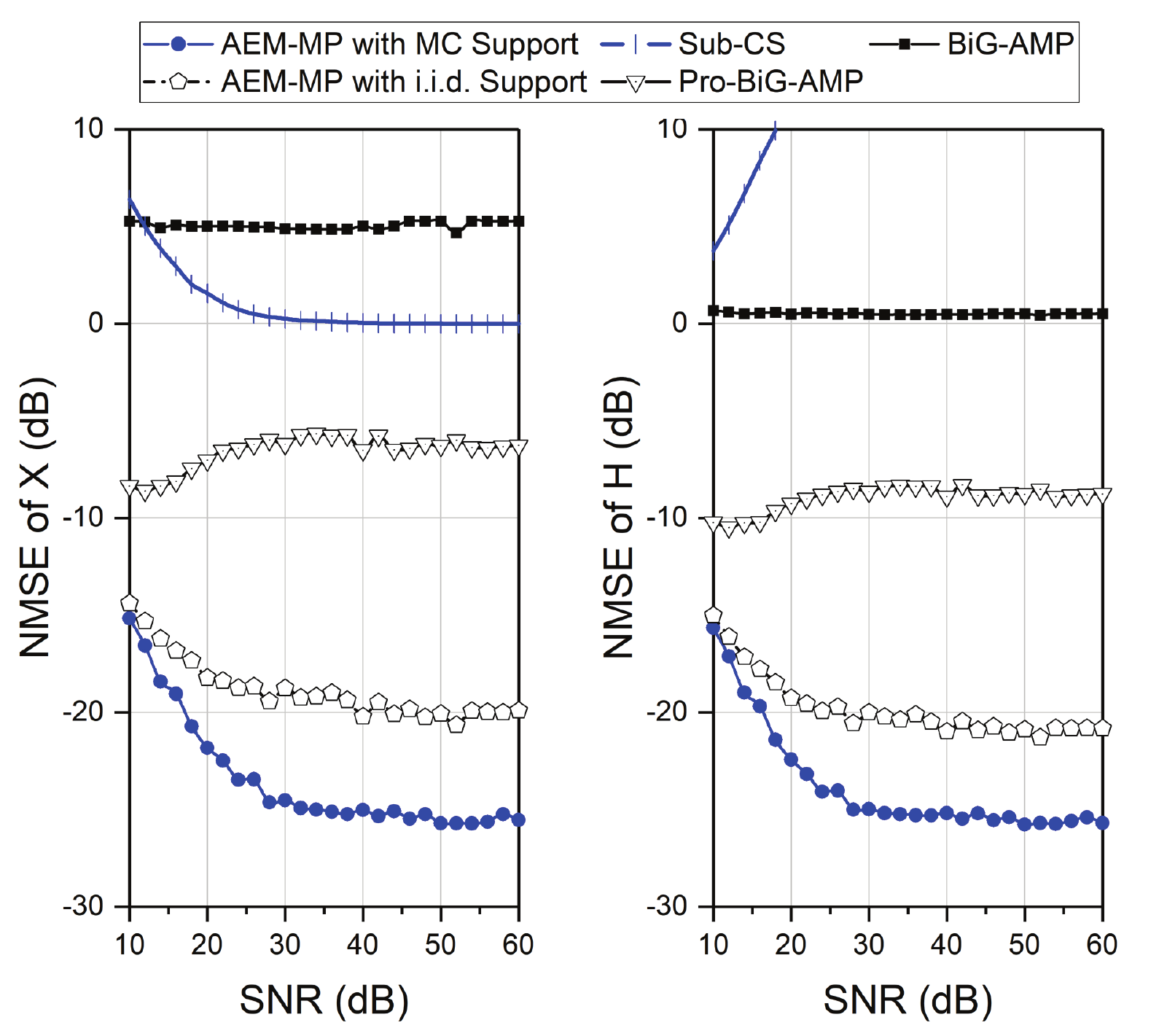}
	\caption{NMSE versus SNR under ULA with true AoAs unknown to the BS, where $N=128$, $T=100$, $K=8$.}
	\label{snr_ula}
\end{figure}
Since the DFT-based methods are applicable to ULAs only, we focus on the simulations with ULAs in this section. The steering vector $\av(\theta)$ is given by \eqref{ULA}, where the inter-antenna spacing is half-wavelength, \ie $d=\varrho/2$. 

First of all, we study the effect of the energy leakage problem on the channel estimation accuracy for different methods. Consider a simple situation where the grid $\varthetav$ corresponding to true AoAs are given by  
\begin{equation}
\sin(\vartheta_l)=\sin(\vartheta^0_l)+\kappa_l, \text{with } \kappa_l \sim U\left[-\frac{1}{L},\frac{1}{L}\right], \forall l \in [L], 
\end{equation}
where $\varthetav^0$ is defined in \eqref{VCR}, and $U([a,b])$ denotes the uniform distribution over $[a,b]$. We further assume that $\{\kappa_l\}_{l=1}^L$ is known at the BS. In other words, the grid $\varthetav$ samples the true AoAs.
Fig. \ref{snr_off} plots the NMSEs of $\Xv$ and $\Hv$ versus different SNR levels with $N\!=\!L\!=\!128$, $T\!=\!100$, $K\!=\!8$, and $\lambda\!=\!0.2$. We see that the performance of all DFT-based methods does not improve as the SNR increases, because the unavoidable energy leakage problem dominates the estimation error in the high SNR regime.  Meanwhile, the AEM-MP algorithm achieves a significant gain with monotonically decreasing NMSE against SNR by avoiding the leakage of energy.
\begin{figure}[!t]
		\centering
		\includegraphics[width=3.2in]{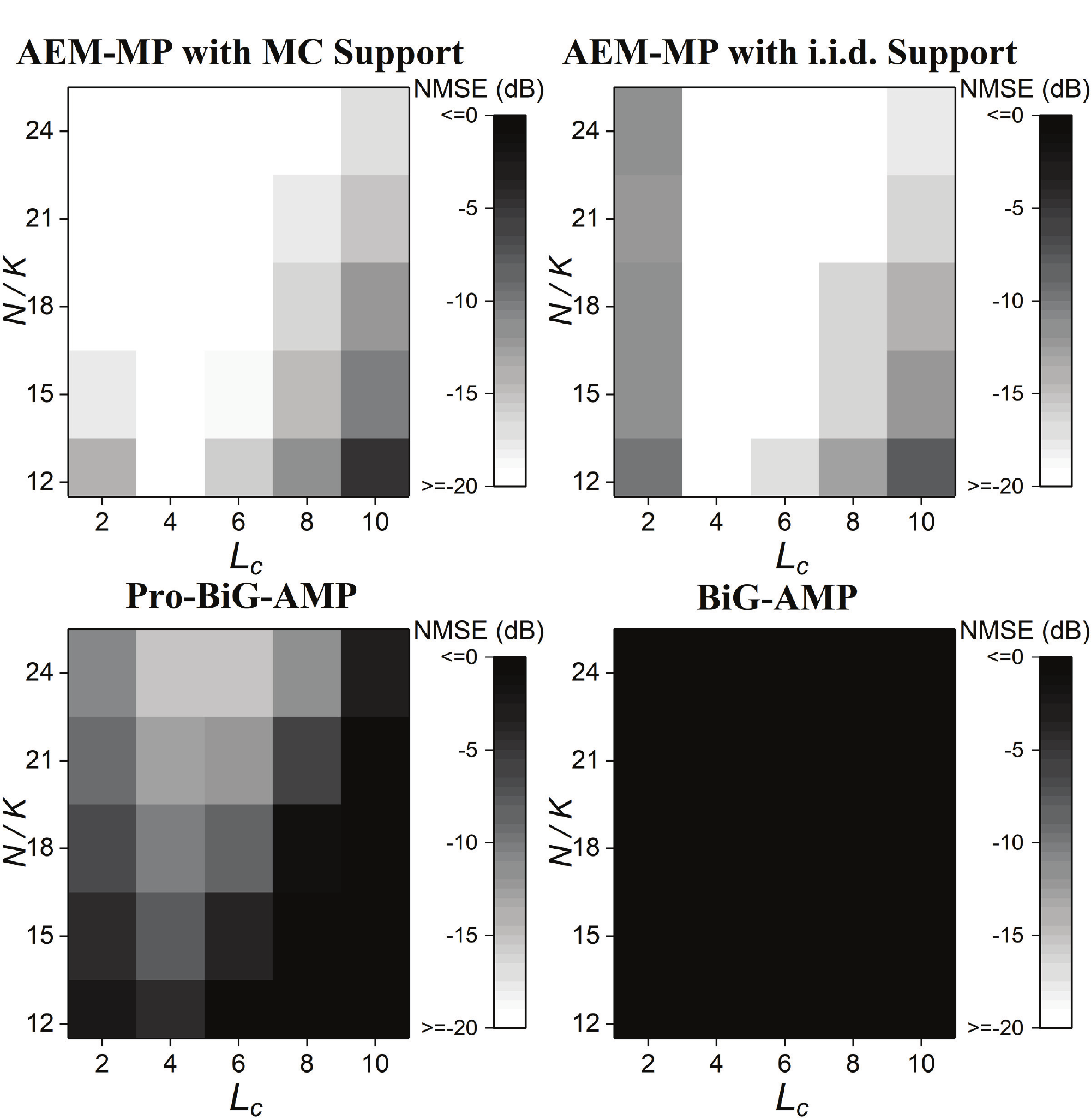}
		\caption{NMSEs of $\Xv$ for different algorithms, where $K\!=\!8$, $T\!=\!100$, and $\text{SNR}\!=\!50\text{  dB}$.}
		\label{phase_ula}
\end{figure}

Now suppose that the AoAs are unknown at the BS. The channel vectors are generated following \eqref{hktilde}. Specifically, we generate the center angle of each scattering cluster uniformly from $\left[-90^{\circ},90^{\circ}\right]$, and the AoA of each subpath $\theta_{k}(i,j)$ concentrates in a $20^{\circ}$ angular spread. Fig. \ref{snr_ula} illustrates the estimation accuracy under various SNRs with  $N\!=\!128$, $T\!=\!100$, $K\!=\!8$, $L_c=3$, and $L_p\!=\!40$. We test the proposed algorithm with Markovian support model \eqref{gc2} and \iid Bernoulli support prior \eqref{gc1}, respectively, with $L$ set to $160$. It can be seen from the figure that 1) accuracies of all DFT-based methods are compromised due to the angle mismatch, which is discussed in Section \ref{DFTdefects}; 
2) our proposed super-resolution estimation algorithms significantly outperform the existing DFT-based methods for all SNR levels, where the improvement is owed to the alleviation of the AoA mismatch by angle tuning; 3) the MC model further improves the estimation accuracy compared to the \iid Bernoulli support as the former captures the burst sparsity of the propagation channels.

The phase transition diagrams are shown in Fig. \ref{phase_ula} for $K=8$, $T=100$, and $\text{SNR}=50\text{ dB}$. Note that if the tuned grid $\varthetav$ matches the true AoAs ideally, $\lambda \propto L_c L_p /L$. Therefore, in simulations we represent the sparsity level of $\Sv$ by $L_c$ and keep $L_p\!=\!40$ and $L\!=\!160$. We demonstrate that 1) as the number of receive antennas $N$ increases, the performance increases for all methods; 2) the relationship between the sparsity level and the estimation error is not monotonic. Similar observations have been previously made in \cite{MIMO_zhangjianwen}.
From the phase diagrams we conclude that our proposed algorithm significantly outperforms the existing methods in all the simulated settings, where the gain mainly comes from the elimination of the angle mismatch.
\begin{figure}[!t]
	\centering
	\includegraphics[width=3.1in]{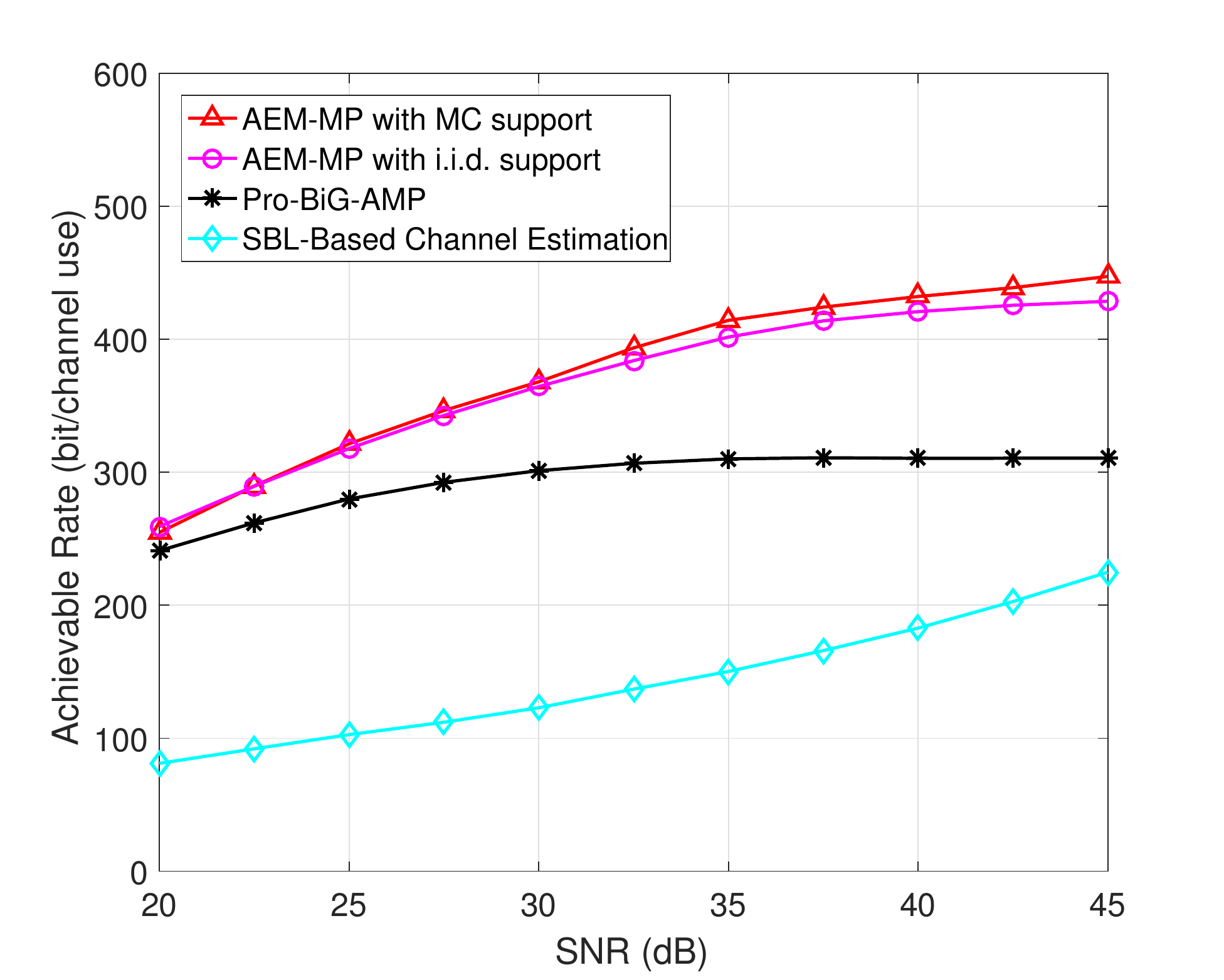}
	\caption{The performance of various schemes under ULA, where $N=640$, $T=120$, $K=40$.}
	\label{ACH_ULA}
\end{figure}

In Fig. \ref{ACH_ULA}, we compare our methods with the SBL-based channel estimation algorithm in terms of the achievable rate. The DFT-based method Pro-BiG-AMP is also included for comparison. 
We set $N=640$, $T=120$, and $K=40$. The grid length $L$ is fixed to $800$ for the non-DFT based estimation methods. All the results are obtained by averaging over $100$ Monte Carlo trials.
It is shown that 1) when SNR is larger than $35\text{ dB}$, the performance of Pro-BiG-AMP does not improve further. The reason is that the DFT basis mismatch and the energy leakage problem dominate the estimation error in the high SNR regime. Similar observations can be found in Fig. \ref{snr_ula}; 2) the achievable rate of the SBL-based training algorithm monotonically increases as SNR increases. However, there exists a significant performance loss compared with our methods due to the estimation error and the rate loss caused by the training overhead; 3) our proposed methods outperform the two baseline schemes. Moreover, the Markovian support model \eqref{gc2} can slightly improve the performance, as it provides more accurate channel-and-signal estimates than the Bernoulli prior \eqref{gc1} does.
\subsection{Blind Channel-and-Signal Estimation with non-ULA}
\label{SIMU2}
As mentioned in Section \ref{DFTdefects}, our proposed algorithms are applicable to general one-dimensional array geometry. First, we take the non-phased antenna array LAA in \eqref{LAA} as an example to demonstrate the capability of the AEM-MP algorithm with non-linear geometry.\footnote{Rigorously speaking, LAA is a two-dimensional array. However, with specific antenna location, its steering vector is related to the azimuth angle only, which is similar to one-dimensional linear arrays.}

\begin{figure}[!t]
	\centering
	\includegraphics[width=3.4in]{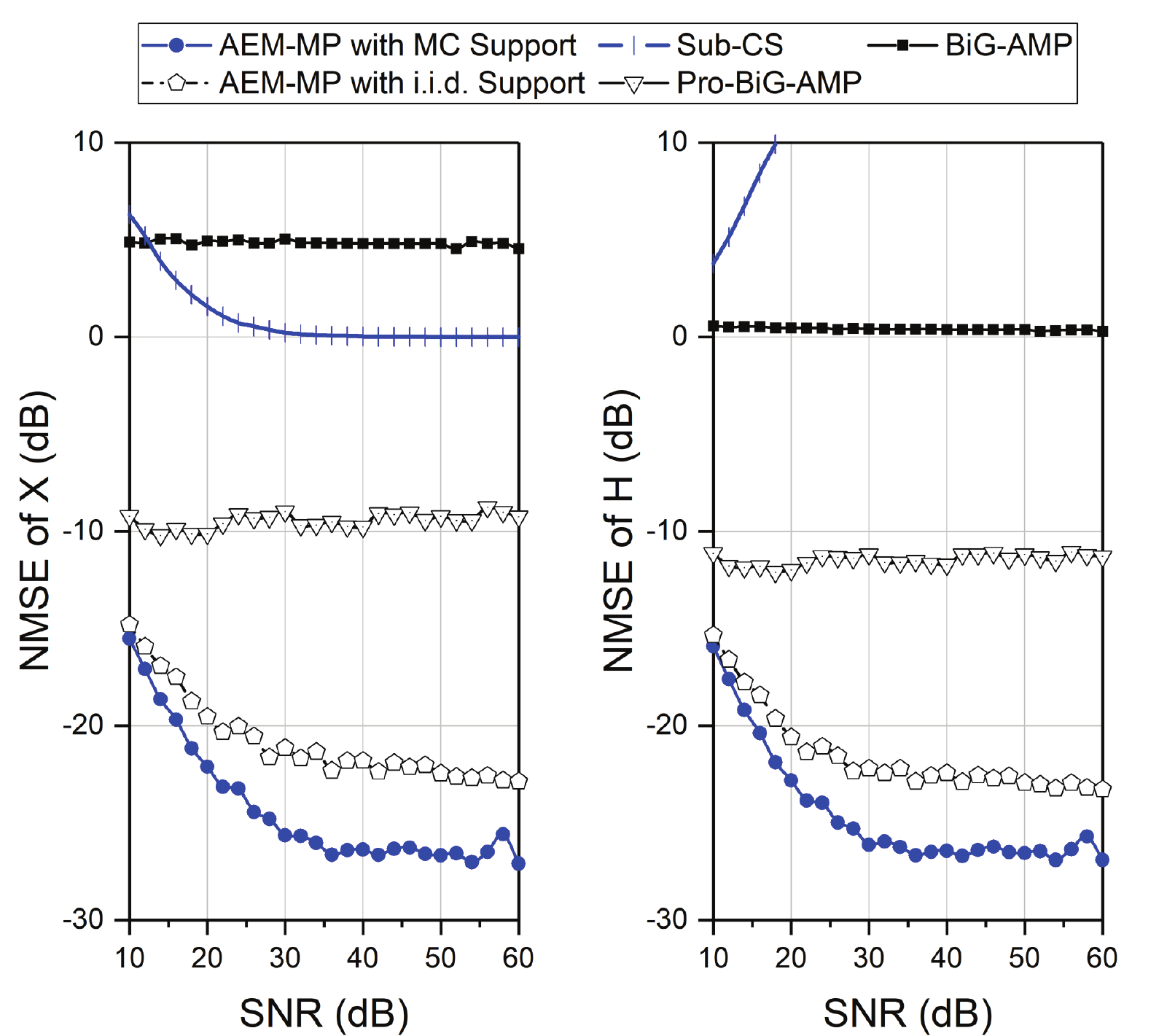}
	\caption{NMSE versus SNR under LAA, where $N=128$, $T=100$, $K=8$.}
	\label{snr_laa}
\end{figure}

\begin{figure}[!tbp]
	\centering
	\includegraphics[width=3.1in]{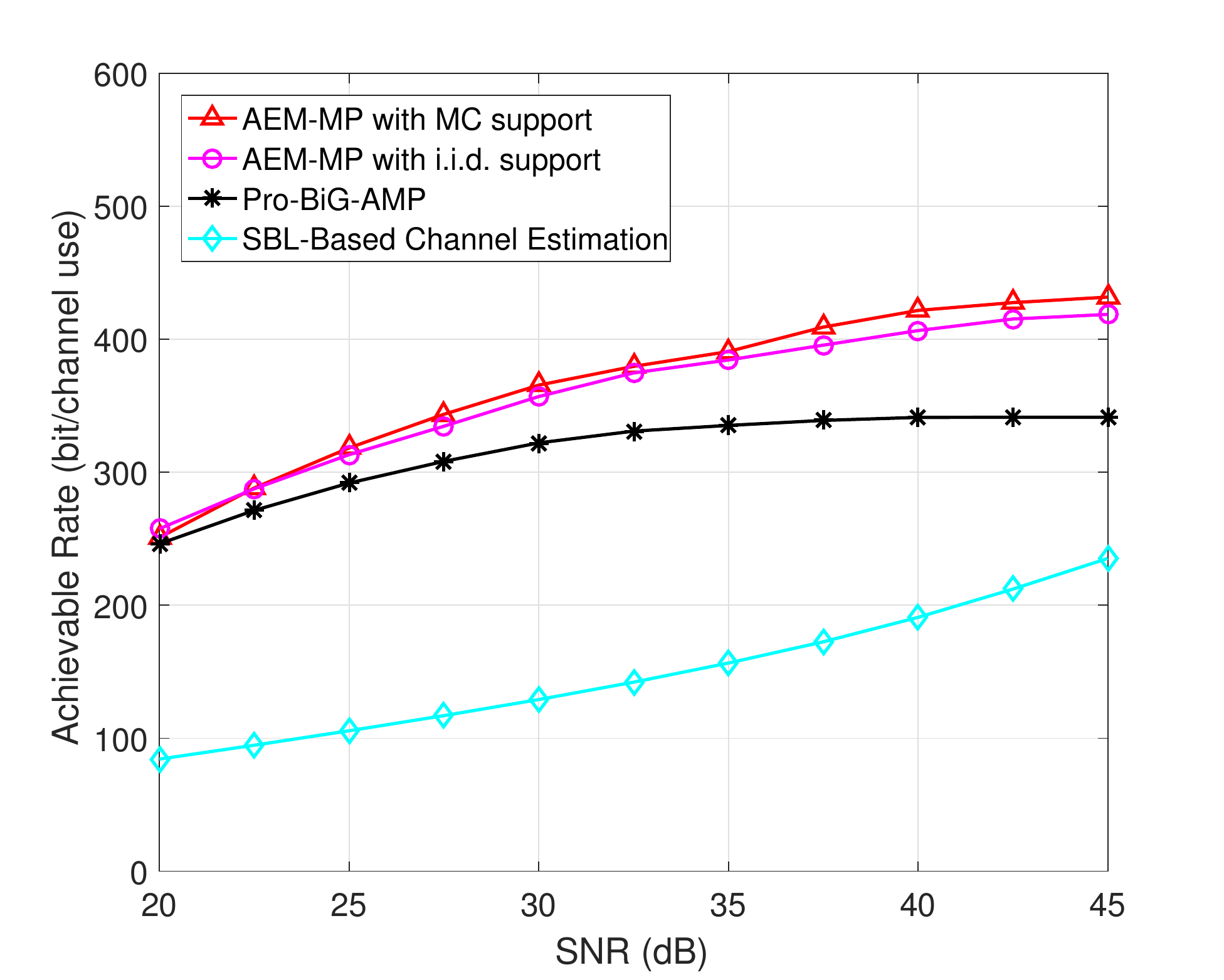}
	\caption{The average achievable rate versus SNR under LAA, where $N=640$, $T=120$, $K=40$.}
	\label{ACH_LAA}
\end{figure}

\begin{figure}[!t]
	\centering
	\includegraphics[width=3.2in]{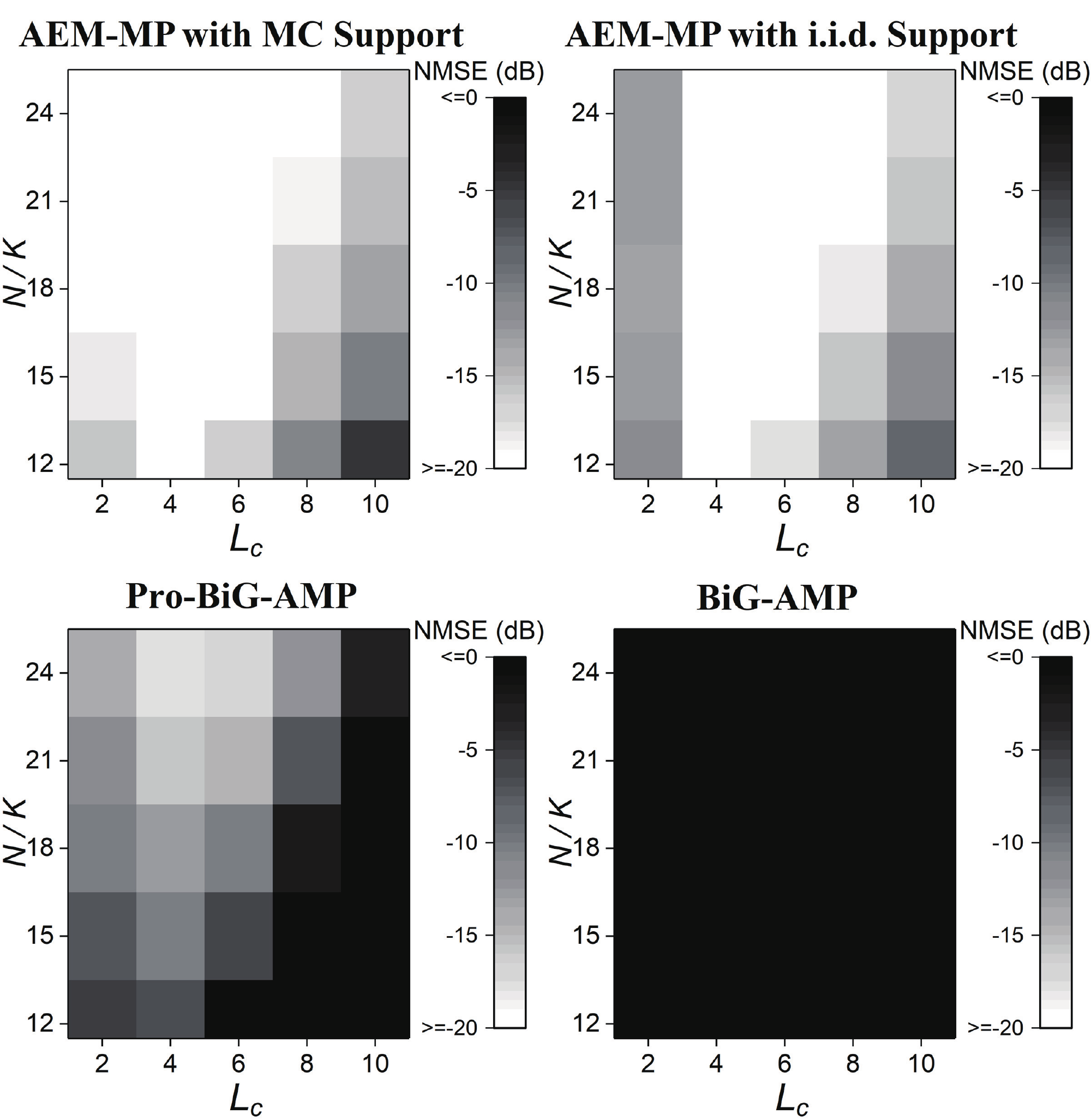}
	\caption{Phase transition diagrams under LAA, where $K\!=\!8$, $T\!=\!100$, and $\text{SNR}\!=\!50\text{ dB}$.}
	\label{phase_laa}
\end{figure}

In Fig. \ref{snr_laa}, we plot the NMSE performance versus SNR under LAA with $N\!=\!128$, $T\!=\!100$, $K\!=\!8$, and $L=160$. The channel is generated similarly to that in Section \ref{SIMU1}, except that the steering vector changes to \eqref{LAA}. We observe similar results as in ULA, where both of the proposed algorithms outperform all the baselines, regardless of the SNR used. Moreover, the achievable rate plots and the phase transition plots are shown in Fig. \ref{ACH_LAA} and Fig. \ref{phase_laa}, respectively. We conclude from the simulations that our algorithms are indeed capable to operate with LAA to produce stable and accurate channel-and-signal estimation.

\begin{figure}[!t]
	\centering
	\includegraphics[width=3.4in]{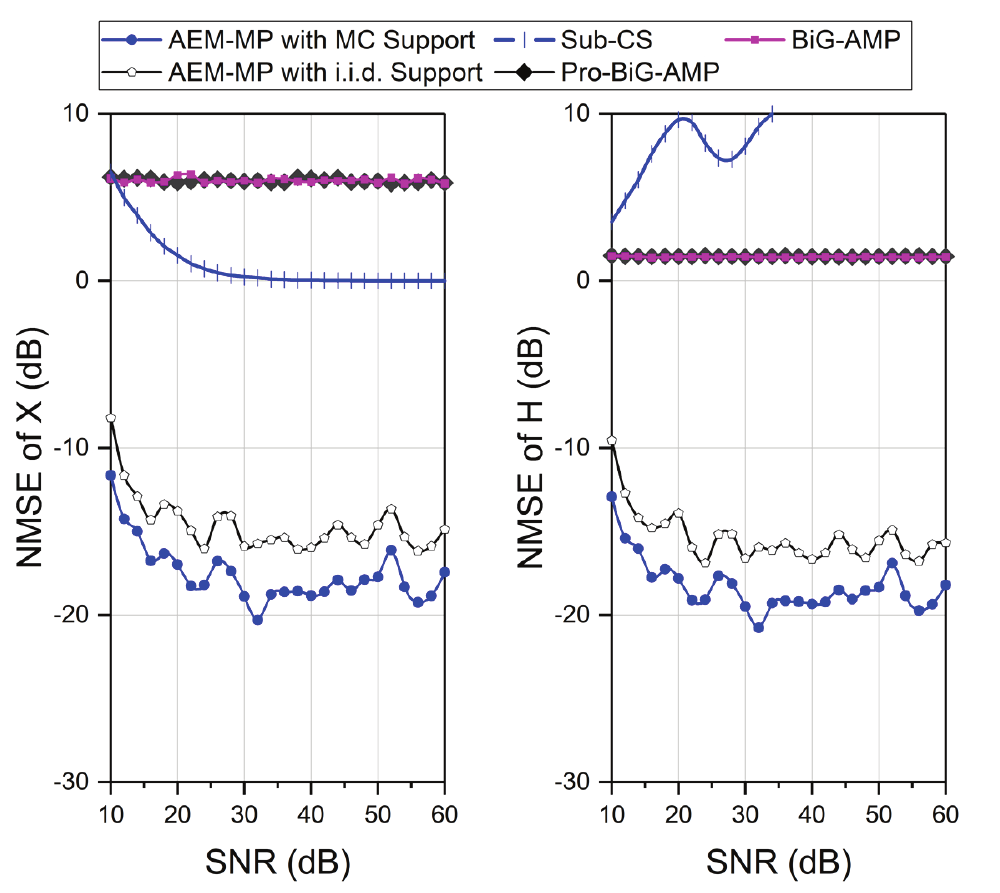}
	\caption{NMSE versus SNR under arbitrary linear arrays, where $N=128$, $T=100$, $K=8$.}
	\label{LA}
\end{figure}

Next, we study the behavior of the blind estimation scheme under arbitrary linear arrays. The steering vector $\av(\theta)$ for a general linear array is given by
{\begin{align}
          \label{LINEARARRAY}
\av(\theta)=\frac{1}{\sqrt{N}}\left[1,e^{-j\frac{2\pi}{\varrho}d_2\sin(\theta)},\cdots,e^{-j\frac{2\pi}{\varrho}d_{N}\sin(\theta)}\right]^T,
\end{align} 
where $d_i$ denotes the distance between the $i$-th and the first antennas.} We  plot the NMSEs for various schemes under arbitrary linear arrays in  Fig. \ref{LA} with $N\!=\!128$, $T\!=\!100$, $K\!=\!8$, and $L=160$. The channel is generated with the steering vector given by \eqref{LINEARARRAY}, where $d_i$ is uniformly generated from $[d_{i\!-\!1}\!+0.4\varrho,d_{i\!-\!1}\!+0.5\varrho]$.
On one hand, the NMSEs of $\Xv$ for all DFT-based methods are greater than $1$. The reason is that under non-uniform linear array geometry, the mismatch between the DFT matrix and the true array response becomes serious.
On the other hand, both of our proposed methods outperform the baseline schemes with significantly higher accuracy, which verifies the efficiency and robustness of our methods. Moreover, Markovian support prior \eqref{gc2} provides better accuracy than \iid Bernoulli prior \eqref{gc1}, which is similar to the ULA case.

{Finally, we investigate the performance of the proposed methods under two-dimensional array geometry. We assume that an $N_1\times N_2$ uniform rectangular array (URA) is deployed at the BS with $N_1\cdot N_2=N$. The steering vector $\av(\theta,\psi)\in \Complex^{N\times 1}$ is given by
	\begin{align}\label{temp61}
\av(\theta,\psi)=\av_v(\theta,\psi)\otimes \av_h(\theta,\psi),
\end{align}
	where $\psi$ denotes the elevation AoA; $\otimes$ denotes the Kronecker product; $\av_h(\theta,\psi)\in \Complex^{N_1\times 1}$ and $\av_v(\theta,\psi)\in\Complex^{N_2\times 1}$ are the steering vectors in the horizontal and vertical directions, respectively. Specifically, they are given by
	\begin{subequations}
	\begin{align}
[\av_h(\theta,\psi)]_n\!=&\frac{1}{\sqrt{N_1}}e^{-j\frac{2\pi d}{\varrho}(n\!-\!1)\cos(\psi)\sin(\theta)}, \forall n \in [N_1],\\
	[\av_v(\theta,\psi)]_n\!=&\frac{1}{\sqrt{N_2}}e^{j\frac{2\pi d}{\varrho}(n\!-\!1)\cos(\psi)\cos(\theta)}, \forall n \in [N_2].
 	\end{align}
	Fig. \ref{URA} shows the NMSE performance over $100$ Monte Carlo channel realizations with $N_1\!=N_2\!=\!30$, $T\!=\!100$, $K\!=\!8$, and $L=160$. The channel is generated with the steering vector \eqref{temp61} and $d=\varrho/2$. The center azimuth and elevation AoAs of the clusters are uniformly drawn from $[-180^\circ,180^\circ]$ and $[-25^\circ,25^\circ]$, respectively. It verifies that our proposed methods indeed cope with the two-dimensional array, whereas the baselines all have poor NMSE performance because of the severe energy leakage.
	
\end{subequations}
	\begin{figure}[!t]
		\centering
		\includegraphics[width=3.4in]{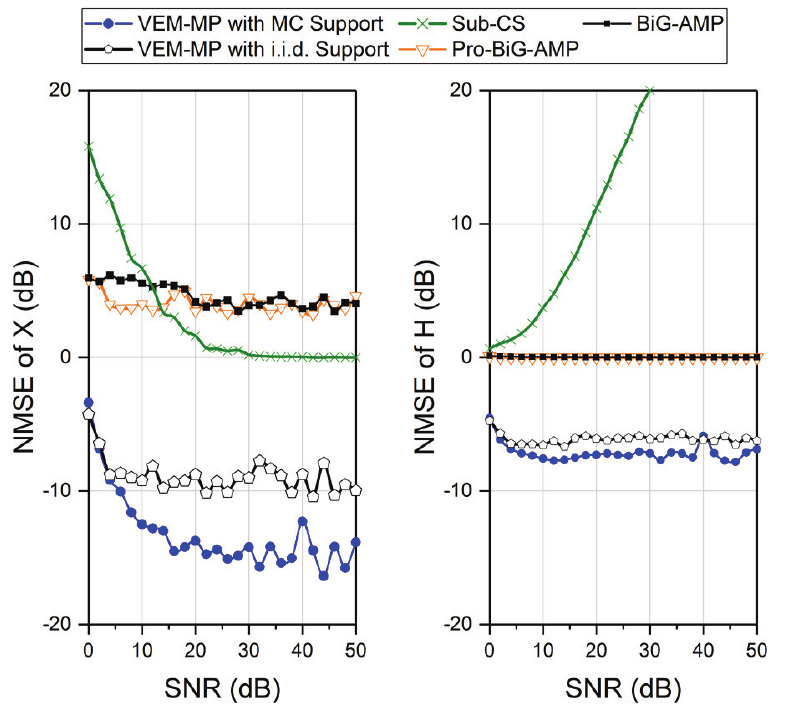}
		\caption{NMSE versus SNR under URA, where $N_1\!=N_2\!=\!30$, $T\!=\!100$, and $K\!=\!8$.}
		\label{URA}
	\end{figure}
}

\section{Conclusions}
\label{conclusions}
In this paper, we studied the channel-and-signal estimation problem in massive MIMO systems. We investigated the angle mismatch and performance loss problem in all existing DFT-based methods. To tackle this challenge, we presented a non-uniform sampling grid channel representation with the Markovian support. Then, we proposed a novel blind channel-and-signal estimation algorithm on top of this model. Furthermore, we developed the message passing algorithm for the marginal computation in the implementation of our proposed algorithm. Finally, numerical results demonstrate the significant estimation accuracy improvement of our proposed algorithm compared to the existing schemes.
\appendices
%
%
\section{\label{mpproof}}
We derive the messages in the large-system limit, \ie $N,K,L,T \to \infty$ with fixed ratios for $L/N$, $K/N$, and $T/N$. Without loss of generality, we assume that $x_{kt}$ scales as $O(1/\sqrt{T})$ and $A_{nl}$ scales as $O(1/\sqrt{N})$, other quantities scale as $O(1)$.
\subsubsection{Derivation of the messages within the AMF structure}

Based on the following result, we can obtain a tractable approximation for $p^i_{w_{lt}}(w_{lt})$ in \eqref{pw}.
\fact{(cf. \cite[Appendix]{MP_ARM}) Under the large-system limit, \ie $N,K,L,T \to \infty$ with fixed ratios, we can approximate $p^i_{w_{lt}}(w_{lt})$ by the Gaussian distribution, with mean $\ph_{lt}(i)$ and variance $v^p_{lt}(i)$ given by
\begin{eqnarray*}
\ph_{lt}(i)\!\!&=&\!\!\sum_{k=1}^K \sh_{lk,t}(i)\xh_{kt,l}(i),\\
v^p_{lt}(i)\!\!&=&\!\!\sum_{k=1}^K \abs{\sh_{lk,t}(i)}^2v^x_{kt,l}(i)\!+\!\sum_{k=1}^K v^s_{lk,t}(i)  \abs{\xh_{kt,l}(i)}^2\!\nonumber\\
&&+\!\sum_{k=1}^K v^s_{lk,t}(i)v^x_{kt,l}(i).
\end{eqnarray*}
\label{le1}
}
Justifications for Fact \ref{le1} can be found in eqs. (78)--(81) of \cite{MP_ARM}, which is based on the argument of eqs. (50)--(52) in \cite{MP_BIGAMP3}.

Then, we employ the methodology in the derivation of AMP \cite{MP_GAMP} to approximate \eqref{ztow}--\eqref{wtoz}. Based on the central limit theorem argument, we approximate $\prod_{l^\prime \! \neq l} \Delta^i_{f_{l^\prime \! t}\to z_{nt}}$ in \eqref{ztow} by the Gaussian distribution. Substituting the Gaussian term into \eqref{ztow}--\eqref{wtoz}, we have  $\Delta_{f_{lt}\to z_{nt}}^i(w_{lt})\approx \CN(w_{lt};\wh_{lt,n}(i),v^w_{lt,n}(i))$, where the mean and variance are given by
\begin{subequations}
	\begin{eqnarray}
	\wh_{lt,n}(i\!+\!1)\!\!&\approx&\!\!\wh_{lt}(i\!+\!1)\!-\! v^w_{lt}(i\!+\!1)A_{nl}\betah_{nt}(i) ,\label{vwh_nnn}\\
	v^w_{lt,n}(i\!+\!1)\!\!&\approx&\!\!v^w_{lt}(i\!+\!1),\label{wh_nnn}
	\end{eqnarray}
\end{subequations}
\begin{subequations}
with
\begin{eqnarray}
v^w_{lt}(i\!+\!1)\!\!&=&\!\!\frac{v^p_{lt}(i)v^\zeta_{lt}(i)}{v^p_{lt}(i)\!+\!v^\zeta_{lt}(i)},\label{vwh_nn}\\
\wh_{lt}(i\!+\!1)\!\!&=&\!\!v^w_{lt}(i\!+\!1)\left( \frac{\zetah_{lt}(i)}{v^\zeta_{lt}(i)}\!+\!\frac{\ph_{lt}(i)}{v^p_{lt}(i)}\right) ,\label{wh_nn}
\end{eqnarray}
\end{subequations}
where the auxiliary variables are given by
\begin{subequations}
\begin{eqnarray}
v^u_{nt}(i)\!\!&=&\!\! \sum_{l=1}^L \abs{A_{nl}}^2 v_{lt}^w(i),\label{vuh_nn}\\
\uh_{nt}(i)\!\!&=&\!\!\sum_{l=1}^L A_{nl}\wh_{lt}(i)\!-\!v^u_{nt}(i)\betah_{nt}(i-1),\label{uh_nn}\\
v^z_{nt}(i)\!\!&=&\!\!\frac{v^u_{nt}(i)\sigma^{2}}{v^u_{nt}(i)\!+\!\sigma^{2}},\label{vzh}\\
\zh_{nt}(i)\!\!&=&\!\!\frac{\uh_{nt}(i)\sigma^{2}\!+\!y_{nt}v^u_{nt}(i)}{v^u_{nt}(i)\!+\!\sigma^{2}}, \label{zh}\\
v^\beta_{nt}(i)\!\!&=&\!\!\frac{v^u_{nt}(i)\!-\!v^z_{nt}(i)}{(v^u_{nt}(i))^2},\label{vtauh}\\
\betah_{nt}(i)\!\!&=&\!\!\frac{\zh_{nt}(i)\!-\!\uh_{nt}(i)}{v^u_{nt}(i)}, \label{tauh}
\end{eqnarray}
\begin{eqnarray}
v^\zeta_{lt}(i)\!\!&=&\!\!\left(\sum_{n=1}^N  \abs{A_{nl}}^2v^\beta_{nt}(i)\right) ^{-1},\label{vzetah}\\
\zetah_{lt}(i)\!\!&=&\!\!\wh_{lt}(i)\!+\!v^\zeta_{lt}(i)\sum_{n=1}^N A_{nl}^\star \betah_{nt}(i).\label{zetah}
\end{eqnarray}
\end{subequations}
Note that in \eqref{uh_nn}, a first-order correction term is introduced, which is similar to the ``Onsager term" in AMP; see, \eg eq. (2) in \cite{MP_AMP1}.

By \eqref{vzetah} and \eqref{zetah} we have
\begin{eqnarray}
\label{temp6}
\prod_{n=1}^N \Delta_{z_{nt}\to f_{lt}}^i(w_{lt})\approx \CN(w_{lt};\zetah_{lt}(i),v^\zeta_{lt}(i)).
\end{eqnarray}
Substituting \eqref{temp6} into \eqref{wtox} and \eqref{wtos}, we find that the forms of \eqref{wtox} and \eqref{wtos} match that in eq. (13) of \cite{MP_BIGAMP1}. Following Section II-D to Section II-F in \cite{MP_BIGAMP1}, we compute $\ph_{lt}(i)$ and $v^p_{lt}(i)$ and close the loop as
\begin{subequations}
\begin{align}
v^p_{lt}(i)=&\sum_{k=1}^K\left(  \abs{\sh_{lk}(i)}^2v^x_{kt}(i)\!+\!v^s_{lk}(i) \abs{\xh_{kt}(i)}^2\!+\!v^s_{lk}(i)v^x_{kt}(i)\right) ,  \label{vph2}\\
\ph_{lt}(i)=&\sum_{k=1}^K \sh_{lk}(i)\xh_{kt}(i) \nonumber\\
& \!-\!\gammah_{lt}(i\!-\!1)\sum_{k=1}^K\left( \abs{\sh_{lk}(i)}^2v^x_{kt}(i)\!+\!v^s_{lk}(i) \abs{\xh_{kt}(i)}^2\right), \label{ph2}\\
v^\gamma_{lt}(i)=&\left(v^p_{lt}(i)\!+\!v^\zeta_{lt}(i)\right)^{-1}, \label{valphah2}\\
\gammah_{lt}(i)=&\frac{\zetah_{lt}(i)\!-\!\ph_{lt}(i)}{v^p_{lt}(i)\!+\!v^\zeta_{lt}(i)}, \label{alphah2}\\
v^r_{kt}(i)=&\left(\sum_{l=1}^L \abs{\sh_{lk}(i)}^2v^\gamma_{lt}(i)\right)^{-1},\label{vrh2}\\
\rh_{kt}(i)=& \left(1\!-\!v^r_{kt}(i)\sum_{l=1}^L v^s_{lk}(i)v^\gamma_{lt}(i)\right)\xh_{kt}(i)\\
&\!+\!v^r_{kt}(i)\sum_{l=1}^L \sh_{lk}^{\star}(i)\gammah_{lt}(i),\label{rh2}\\
v^q_{lk}(i)=&\left(\sum_{t=1}^T \abs{\xh_{kt}(i)}^2v^\gamma_{lt}(i)\right)^{-1},\label{vqh2}\\
\qh_{lk}(i)=&\left(\!1\!-\!v^q_{lk}(i)\sum_{t=1}^T v^x_{kt}(i)v^\gamma_{lt}(i)\!\right)\sh_{lk}(i)\nonumber\\
&\!+\!v^q_{lk}(i)\!\sum_{t=1}^T \xh_{kt}^{\star}(i)\gammah_{lt}(i),\label{qh2}
\end{align}
\end{subequations}
where 
\begin{subequations}
\begin{eqnarray}
\xh_{kt}(i\!+\!1)\!\!&=&\!\!\int_{x_{kt}} x_{kt} \Delta_{x_{kt}}^i(x_{kt}),\label{xposta} \\
v^x_{kt}(i\!+\!1)\!\!&=&\!\!\int_{x_{kt}} x^2_{kt} \Delta_{x_{kt}}^i(x_{kt})-\abs{\xh_{kt}(i\!+\!1)}^2,\\
\sh_{lk}(i\!+\!1)\!\!&=&\!\!\int_{s_{lk}} s_{lk} \Delta_{s_{lk}}^i(s_{lk}),
\end{eqnarray}
\begin{eqnarray}
v^s_{lk}(i\!+\!1)\!\!&=&\!\!\int_{s_{lk}} s_{lk}^2 \Delta_{s_{lk}}^i(s_{lk})-\abs{\sh_{lk}(i\!+\!1)}^2.\label{sposta}
\end{eqnarray}
\end{subequations}
\begin{figure*}[!t]
	\begin{align}
	\lambda^{\text{fwd}}_{lk}(i)\!&=\!\begin{cases}
	\lambda,\!\!&\!\!l\!=\!1,\\
	\frac{p_{10}(1\!-\!\lambda^{\text{fwd}}_{l\!-\!1,k}(i))(1\!-\!\pi^{\text{out}}_{l\!-\!1,k}(i))\!+(1\!-\!p_{01})\!\lambda^{\text{fwd}}_{l\!-\!1,k}(i)\pi^{\text{out}}_{l\!-\!1,k}(i)}{(1\!-\!\lambda^{\text{fwd}}_{l\!-\!1,k}(i))(1\!-\!\pi^{\text{out}}_{l\!-\!1,k}(i))\!+\!\lambda^{\text{fwd}}_{l\!-\!1,k}(i)\pi^{\text{out}}_{l\!-\!1,k}(i)},\!\!&\!\!l\!\geq\!2.
	\end{cases}
	\label{fwd1}\\
	\lambda^{\text{bwd}}_{lk}(i)\!&=\!\begin{cases}
	\frac{1}{2},\!\!&\!\!l\!=\!L,\\
	\frac{p_{01}(1\!-\!\lambda^{\text{bwd}}_{l\!+\!1,k}(i))(1\!-\!\pi^{\text{out}}_{l\!+\!1,k}(i))\!+(1\!-\!p_{01})\!\lambda^{\text{bwd}}_{l\!+\!1,k}(i)\pi^{\text{out}}_{l\!+\!1,k}(i)}{(1\!-\!p_{10}\!+\!p_{01})(1\!-\!\lambda^{\text{bwd}}_{l\!+\!1,k}(i))(1\!-\!\pi^{\text{out}}_{l\!+\!1,k}(i))\!+\!(1\!-\!p_{01}\!+\!p_{10})\lambda^{\text{bwd}}_{l\!+\!1,k}(i)\pi^{\text{out}}_{l\!+\!1,k}(i)},\!\!&\!\!l\!<\!L.
	\end{cases}
	\label{bwd}\\
	\omega_{lk}(i\!+\!1)\!&=\!\frac{\lambda^{\text{fwd}}_{lk}(i)\lambda^{\text{bwd}}_{lk}(i)\pi^{\text{out}}_{lk}(i)}{\lambda^{\text{fwd}}_{lk}(i)\lambda^{\text{bwd}}_{lk}(i)\pi^{\text{out}}_{lk}(i)\!+\!(1\!-\!\lambda^{\text{fwd}}_{lk}(i))(1\!-\!\lambda^{\text{bwd}}_{lk}(i))(1\!-\!\pi^{\text{out}}_{lk}(i))}.\label{temp64}
	\end{align}
	\hrulefill
\end{figure*}
\subsubsection{Derivation of $\Delta_{j_{lk}\to c_{lk}}^i$ and $\Delta_{j_{l+1,k}\to c_{lk}}^i$}
Plugging the values of $\Delta_{j_{l\!-\!1,k}\!\to\! c_{l\!-\!1,k}}^i$ and $\Delta_{h_{l\!-\!1,k}\!\to\!c_{l\!-\!1,k}}^i$ into \eqref{jtoc}, we obtain \eqref{fwd1} for $\Delta_{j_{lk}\to c_{lk}}^i(c_{lk}\!=\!1)\triangleq\lambda^{\text{fwd}}_{lk}(i)$,
Similarly, we compute $\Delta_{j_{l+1,k}\to c_{lk}}^i(c_{lk}=1)\triangleq \lambda^{\text{bwd}}_{lk}(i)$ as \eqref{bwd}.
\subsubsection{Derivation of $\Delta_{h_{lk}\to c_{lk}}^i$, $\Delta_{c_{lk}\to h_{lk}}^i$ and $\Delta_{c_{lk}}^{i\!+\!1}$} We finish the whole derivation by computing $\Delta_{h_{lk}\to c_{lk}}^i$, $\Delta_{c_{lk}\to h_{lk}}^i$, and $\Delta_{c_{lk}}^{i\!+\!1}$, while the remaining results in Table \ref{table2} can be obtained simply by the sum-product rules. Note that
\begin{align}
\Delta_{h_{lk}\to c_{lk}}^i(c_{lk})\!&\propto\! \int_{s_{lk}} \Delta_{s_{lk}\to h_{lk}}^i(s_{lk})h_{lk}(s_{lk},c_{lk})\nonumber\\
\!&\propto\! \pi^{\text{out}}_{lk}(i)\delta\left(c_{lk}\!-\!1 \right)\!+\!(1\!-\!\pi^{\text{out}}_{lk}(i))\delta\left(c_{lk} \right),
\end{align}
where
\begin{align}
\frac{1}{\pi^{\text{out}}_{lk}(i)}&=\frac{\CN(0;\qh_{lk}(i),v^q_{lk}(i))}{\int_{s_{lk}}\CN(s_{lk},\qh_{lk}(i),v^q_{lk}(i))\CN(s_{lk};0,\varphi_k)}\!+\!1\nonumber\\
&=1+\frac{(\varphi_k+v^q_{lk})}{v^q_{lk}}\exp{\left(\frac{\abs{\qh_{lk}}^2}{v^q_{lk}+\varphi_k}-\frac{\abs{\qh_{lk}}^2}{v^q_{lk}}\right)}.\label{piout}
\end{align}
Similarly, for $\Delta_{c_{lk}\to h_{lk}}^i(c_{lk}=1)\triangleq \pi^{\text{in}}_{lk}(i)$,
\begin{align}
\pi^{\text{in}}_{lk}(i)=\frac{\lambda^{\text{fwd}}_{lk}(i)\lambda^{\text{bwd}}_{lk}(i)}{\lambda^{\text{fwd}}_{lk}(i)\lambda^{\text{bwd}}_{lk}(i)\!+\!(1\!-\!\lambda^{\text{fwd}}_{lk}(i))(1\!-\!\lambda^{\text{bwd}}_{lk}(i))}.\label{piin}
\end{align}
From \eqref{cmarg}, we have \eqref{temp64} for $\Delta_{ c_{lk}}^{i\!+\!1}(c_{lk}\!=\!1)\triangleq \omega_{lk}(i\!+\!1)$.

\section{Derivation of \eqref{temp13}}
\label{pro4}
We omit the superscript $(\cdot)^{(j)}$ for the ease of notation. From \eqref{up2} and by ignoring the irrelevant terms, we obtain
\begin{equation}
\begin{aligned}
\label{temp20}
\vartheta^{j+1}_l&=	\argmax_{\vartheta_l}	E\Big[\sum_{n=1}^N \sum_{t=1}^T \ln \CN(y_{nt}; z_{nt},\sigma^2)\Big]\\
&=	\argmax_{\vartheta_l} -\sigma^{-2} \norm{\Yv-\Av\Wvh}_F^2 -\sigma^{-2}\sum_{t=1}^T\sum_{l=1}^L v^w_{lt} \norm{\av_l}^2_2,
\end{aligned}
\end{equation}
where
\begin{align}
&\frac{\partial \norm{\Yv\!-\!\Av\Wvh}_F^2}{\partial \vartheta_l}\!=2\!\Re\! \left((\av^\prime_l)^H\!\left(\av_l\sum_{t=1}^T \abs{\wh_{lt}}^2\!-\!\sum_{t=1}^T \wh_{lt}^\star \yv_{t\!-\!l}\right)\right),\label{temp11}
\end{align}
\begin{align}
&\frac{\partial \sum_{t,l} v^w_{lt} \norm{\av_l}^2_2}{\partial \vartheta_l}\!=2\!\Re\! \left(\left(\sum_{t=1}^Tv^w_{lt}\right)(\av^\prime_l)^H \av_l\right)\label{temp22}.
\end{align}
Combining \eqref{temp11} and \eqref{temp22} completes the derivation.
\section{Derivation of \eqref{sigmaupdate}--\eqref{lambdaupdate}}
\label{pro5}
Ignoring the irrelevant terms in \eqref{up1}, we obtain
\begin{align}
\label{temp51}
(\sigma^2)^{(j+1)}\!=\!\argmax_{\sigma^2 > 0}\sum_{n=1}^N \sum_{t=1}^T \int_{z_{nt}}\!\!p(z_{nt}|\Yv;\Psi^{(j)})\!\ln p(y_{nt}|z_{nt};\sigma^2).
\end{align}
Taking the derivative of \eqref{temp51} w.r.t. $\sigma^2$, we have 
\begin{align}
\label{temp52}
\sum_{n=1}^N \sum_{t=1}^T \int_{z_{nt}}p(z_{nt}|\Yv;\Psi^{(j)})\frac{\partial}{\partial (\sigma^2)}\ln p(y_{nt}|z_{nt};\sigma^2)\nonumber\\
=\sum_{n=1}^N \sum_{t=1}^T \int_{z_{nt}}p(z_{nt}|\Yv;\Psi^{(j)})\left(\frac{\abs{y_{nt}-z_{nt}}^2}{(\sigma^2)^2}\!-\!\sigma^{-2} \right).
\end{align}
Setting R.H.S. of \eqref{temp52} to $0$, we reach \eqref{sigmaupdate}.
Similarly, for \eqref{up3},
\begin{align}
\label{temp53}
\varphi_k^{(j+1)}=&\argmax_{\varphi_k >0 }\sum_{l=1}^L \int_{s_{lk}}p(s_{lk}|\Yv;\Psi^{(j)})\ln p_{s_{lk}}(s_{lk};\varphi_k),
\end{align}
where $p_{s_{lk}}(s_{lk};\varphi_k)$ is given in \eqref{gs2}.
Besides, we have
\begin{align}
&p(s_{lk}|\Yv;\Psi^{(j)})\approx \Delta_{s_{lk}} (s_{lk})\nonumber\\
&=\! \frac{\CN(s_{lk};\qh_{lk},v^q_{lk})\left( \pi^{\text{in}}_{lk}\CN(s_{lk};0,\varphi_k^{(j)})\!+\!(1-\pi^{\text{in}}_{lk})\delta(s_{lk})\right) }{\int_{s_{lk}}\!\CN(s_{lk};\qh_{lk},v^q_{lk})\!\left( \pi^{\text{in}}_{lk}\CN(s_{lk};0,\varphi_k^{(j)})\!+\!(1\!-\!\pi^{\text{in}}_{lk})\delta(s_{lk})\right) }\nonumber\\
&= \eta_{lk}^{(j)}\CN(s_{lk};\chi_{lk}^{(j)},\nu_{lk}^{(j)})\!+\!(1\!-\!\eta_{lk}^{(j)})\delta(s_{lk}),\label{temp54}
\end{align}
where we omit the superscript $(\cdot)^{(j)}$ in $\qh_{lk}$, $v^q_{lk}$, $\pi^{\text{in}}_{lk}$, and $\pi^{\text{out}}_{lk}$  for brevity and $\eta_{lk}^{(j)}$, $\nu_{lk}^{(j)}$, and $ \chi_{lk}^{(j)}$ are given by \eqref{temp55}--\eqref{temp57}.
Following the derivation of eq. (46) in \cite{MP_BGEM}, we compute the derivative of the objective in \eqref{temp53} and update $\varphi_{k}$ as \eqref{varphiupdate}.

Taking the derivative of \eqref{up4} w.r.t. $p_{01}$ and ignoring the irrelevant terms, we obtain
\begin{align}
&\frac{\partial}{\partial p_{01}}\sum_{l=1}^{L-1}\sum_{k=1}^K \E_{c_{lk},c_{l\!+\!1,k}|\Yv}\left[\ln p(c_{lk},c_{l\!+\!1,k})\right]\nonumber\\
&=\sum_{l=1}^{L-1}\sum_{k=1}^K \left( \E_{c_{lk}|\Yv}\left[c_{lk}\right]\!-\!\E_{c_{lk},c_{l\!+\!1,k}|\Yv}\left[c_{lk},c_{l\!+\!1,k}\right]\right) /p_{01}\nonumber\\
&-\sum_{l=1}^{L-1}\sum_{k=1}^K\E_{c_{lk},c_{l\!+\!1,k}|\Yv}\left[c_{lk},c_{l\!+\!1,k}\right]/(1\!-\!p_{01}).\label{temp59}
\end{align}
Setting \eqref{temp59} to zero, we obtain \eqref{p01update}.

Similarly, taking the derivative of \eqref{up5} w.r.t. $\lambda$, we obtain
\begin{align}
&\frac{\partial}{\partial \lambda}\sum_{k=1}^K \E_{c_{1k}|\Yv}\left[\ln p(c_{1k})\right]\nonumber\\
&=\sum_{k=1}^K\frac{\E_{c_{1k}|\Yv}\left[c_{1k}\right]}{\lambda}\!-\!\frac{1\!-\!\E_{c_{1k}|\Yv}\left[c_{1k}\right]}{1\!-\!\lambda}.\label{temp60}
\end{align}
Setting \eqref{temp60} to zero, we have \eqref{lambdaupdate}.
\ifCLASSOPTIONcaptionsoff
\newpage
\fi
\ifhavebib
{

}
\else{
}
\fi

\end{document}